\documentclass[]{interact}

\usepackage{epstopdf}
\usepackage[caption=false]{subfig}

\usepackage[numbers,sort&compress]{natbib}
\bibpunct[, ]{[}{]}{,}{n}{,}{,}
\makeatletter
\def\NAT@def@citea{\def\@citea{\NAT@separator}}
\makeatother

\theoremstyle{plain}

\theoremstyle{definition}

\theoremstyle{remark}

\newcommand{\be}{\begin{equation}}
\newcommand{\ee}{\end{equation}}

\begin{document}
 
\title{\Large A note on fitting a generalized Moody diagram for wall modeled Large Eddy Simulations}
\author{
\name{Charles Meneveau\textsuperscript{a} \thanks{\textsuperscript{a}meneveau@jhu.edu}}
\affil{Department of Mechanical Engineering and Center for Environmental and Applied Fluid Mechanics, Johns Hopkins University, Baltimore MD 21218, USA.}}
\maketitle

\begin{abstract}
Motivated by the needs of wall modeled Large Eddy Simulation (LES), we introduce fits to numerical solutions of the Reynolds Averaged Navier-Stokes equations in their simplest near-wall, boundary layer approximation including a mixing-length model. We formulate the problem such that independent dimensionless variables are those directly available in LES. We provide practical fits for the dependent variable, fits that encompass a smooth transition between the viscous sublayer and inertial logarithmic layer, and then progress first considering moderate pressure gradients as well as roughness effects under the assumption that the mixing-length is not affected by the pressure gradient. An alternative fit based on the empirical wall model  of Nickels (J. Fluid Mech. vol.512, pp. 217-239, 2004) is also provided, taking into account possible effects of pressure gradient on turbulence near-wall structure. We then consider the case of general pressure gradients, both favorable and adverse, up to conditions of separation, for both smooth and rough surfaces.  The proposed fitting functions constitute a generalized Moody chart, comply with analytical solutions valid in various asymptotic regimes, and obviate the need for numerical iterative solution methods or near-wall numerical integration of ordinary differential equations during LES. 
\end{abstract}

\begin{keywords}
Turbulence, Wall Model, Large Eddy Simulations  
\end{keywords}
 
\section{Introduction}

Wall-resolving Large-Eddy-Simulation (LES) of high Reynolds number wall-bounded flows continues to be a challenge due to stringent near wall resolution requirements. A large number of grid points is required to resolve the inner, viscous dominated region, and that number increases rapidly with Reynolds number. Conversely, wall modeled LES exhibits a much weaker dependence on Reynolds number and is therefore a necessary choice when applying LES to high Reynolds number wall-bounded flows.
A variety of wall models have been developed for LES and reviews of many of them can be found in Refs. \cite{piomelli2008wall,piomellibalaras02,larsson2016large}. The most frequently used wall model is the so-called equilibrium wall model. There are three most commonly used approaches to implement the equilibrium wall model, each valid in different Reynolds number ranges and types of surfaces. (a) The rough-wall, high Reynolds number wall model, used e.g. in \cite{moeng84,porte2000scale,bou05}: The approach assumes that the streamwise mean velocity profile $\langle u_s(y)\rangle$ in a direction normal to the surface (coordinate $y$) is given by $\langle u_s(y)\rangle=(u_\tau/\kappa) \log(y/z_0)$, where $z_0$ is the roughness length, $\kappa$ the von Karman constant, and $u_\tau$ the (unknown) friction velocity. Evaluated at a distance $y=\Delta_y$ where the streamwise velocity is known from LES (denoted as $U_{\rm LES} = \langle u_s(\Delta_y)\rangle$) it allows solving for $u_\tau$ as function of $U_{\rm LES}$, $\Delta_y$, $\kappa$ and $z_0$.  The assumption is that $\Delta_y$ falls in the logarithmic layer and that $\kappa$ and $z_0$ are known (e.g. $\kappa=0.4$). (b) The smooth surface case at finite Reynolds number: For flows over smooth surfaces, the equilibrium wall model approach is based on the assumed profile $\langle u_s(y)\rangle=u_\tau[\kappa^{-1} \log(yu_\tau/\nu)+B]$, providing a transcendental equation for $u_\tau$ which must be solved iteratively in a code. Specifically, one solves
$U_{\rm LES} =u_\tau[\kappa{-1} \log(\Delta_y u_\tau/\nu)+B]$ for $u_\tau$, for given $U_{\rm LES}$, $\Delta_y$ and $\nu$ (typical parameter values are $\kappa = 0.4$ and $B=5$). Again, this method assumes $\Delta_y$ falls in the logarithmic layer. If $\Delta_y$ falls in the viscous sublayer (approaching wall resolved LES) one must instead assume a linear profile \cite{yang2015integral}, or one can use a smooth fit to the entire profile 
such as the classic fit by Reichardt (1951) \cite{reichardt1951vollstandige} or the recent work in Refs. \cite{luchini2018structure,gonzalez2018large,adler2020wall} including pressure gradient effects.  Typically the fitted solution is for the velocity profile in inner units, which means that further iterative methods are needed to find the friction velocity numerically. (c) Numerical integration of an ordinary differential equation (ODE method): Typically, if one wishes to ensure a smooth transition between the viscous and log-layer regions, to include additional physical effects, or to apply the approach to other variables such as temperature, a common approach is to use numerical solution of an ODE \cite{larsson2016large}. For the case of an equilibrium layer the ODE to be solved for the streamwise velocity reads
\be 
\frac{d}{dy}\left( (\nu+ \nu_T)\frac{d\langle u_s(y)\rangle}{dy}\right) = 0,
\label{eq:simpleODE}
\ee
subject to boundary conditions $\langle u_s(0)\rangle=0$ and $\langle u_s(\Delta_y)\rangle = U_{\rm LES}$. The turbulent eddy viscosity $\nu_T$ can be prescribed using a mixing length model including a van-Driest damping function. 

It would appear useful to cast this sort of ODE into an appropriate dimensionless form, solve it numerically once and for all, and to provide useful fits to the (inverse) solution that can be applied uniformly to a large number of LES cases. One reason that many researchers opt for numerical solution is that the ODE itself depends upon the unknown dimensional parameter $u_\tau$ via the van-Driest damping function and that when written in inner units as function of $y^+$ the equation must be integrated numerically up to a case-dependent position $y^+=\Delta^+$ which itself depends on the unknown value of $u_\tau$.  In this note, we address this issue by rewriting the equation in a non-standard dimensionless form in terms of two Reynolds numbers that facilitates more general applicability for wall modeling. Another reason researchers opt for numerical solution of the boundary layer equation is that it is then possible to include additional physical effects such as pressure gradient, which we shall address here, or handle other fields such as temperature, which will not be covered. 
 
The aims  of this note are thus rather modest, namely to reformulate Eq. \ref{eq:simpleODE} in such a way as to facilitate numerical integration and fitting of the results in the context of wall-modeled LES (WMLES). Specifically, we fit the {\it inverse} of the solution to the velocity profile, i.e. we will be able to find  $u_\tau = f({\text{known variables}})$ directly using relatively simple function evaluations. We also aim to include pressure gradient effects and to merge the resulting fits smoothly to the equilibrium wall model approach valid for rough-wall, very high Reynolds number wall-bounded flows. This note does not include implementation and applications in LES codes, but documents errors and differences between the proposed fits and the full numerical solution of the corresponding (RANS) ODE. Also, we do not address any of the other fundamental issues underlying wall modeling using the equilibrium wall model, such as the log-layer mismatch and challenges associated with modeling non-equilibrium unsteady terms, issues treated  e.g. in Refs. \cite{kawai2012wall,yang2015integral,bae2019dynamic,lozano2020non}. 

It is hoped that the generalized fits provided (a kind of ``generalized Moody diagram'' for wall modeling in LES) can save computational resources and simplify implementations of equilibrium wall models in LES.   

\section{Friction velocity for turbulent equilibrium flow over a smooth wall}
 
We first focus on the simplest case of wall modeling in which we consider only the streamwise direction (subscripts ``s'') without pressure gradient or other acceleration terms. We assume the streamwise velocity away from the wall is known, and denote it by $U_{\rm LES} = \langle u_s(\Delta_y)\rangle$. The unknown to be determined is the friction velocity 
$u_\tau$, from which the (kinematic) wall stress in the streamwise direction can then be evaluated according to $\tau_w=u_\tau^2$ and oriented according to the usual approaches 
\cite{piomelli2008wall,piomellibalaras02,bou05,larsson2016large}. 
To cast the problem into a dimensionless framework, we now define  two Reynolds numbers:
\be  Re_\Delta = \frac{ U_{\rm LES} \Delta_y }{\nu} \,\,\,\,\, {\rm and}  \,\,\,\, Re_{\tau \Delta} = \frac{ u_\tau \Delta_y }{\nu}.
\label{eq:defreyns}
\ee
In WMLES, $Re_\Delta$ is the known input whereas $Re_{\tau \Delta} = \Delta_y^+$ is the unknown output for which we wish to solve and then obtain $u_\tau$. 

Using the usual mixing length model, integrating  Eq. \ref{eq:simpleODE} once and using the fact that the stress tends to $u^2_{\tau}$ as $y \to 0$ we have
\be
\left(\nu+[D(y) \, \kappa \, y\,]^2 \left| \frac{du}{dy} \right| \right) \frac{du}{dy} =  u^2_{\tau},
\ee
where for notational simplicity henceforth we set $u=\langle u_s\rangle$. The traditional van Driest damping function is included: $D(y) = [1-\exp(-y^+/A^+)]$ with $y^+ = (y/\Delta_y) Re_{\tau\Delta}$, and $A^+=25$ is a commonly used  value. This formulation assumes that $\Delta_y$ is sufficiently small so as to not fall into the outer wake region of boundary layers. In WMLES, this condition is typically met as long as more than $O(10)$ grid-points are used to resolve the boundary layer region. In the remainder of this note, we will continue making this assumption. 
 
We first develop a numerical integration by recasting this equation in terms of dimensionless variables that can be expressed in terms of the dimensional parameters known in LES (besides $U_{\rm LES}$), namely $\Delta_y$ and $\nu$:
\be
y'=\frac{y}{\Delta_y},~~~~~ \hat{u}(y') = \frac{u(y) \Delta_y}{\nu} \, .
\ee
The equation then reads as follows:
\be 
\frac{d\hat u}{dy'} + [D(y')\,\kappa \, y'\,]^2 \left( \frac{d\hat u}{dy'} \right)^2 = Re_{\tau \Delta}^2
\ee
(for now we assume a monotonic profile, where $du/dy$ does not change sign). Solving the quadratic equation \cite{hinze59} casts it into a simple first-order ODE for $\hat u(y')$:
\be 
\frac{d\hat u}{dy'} =  \frac{1}{2 [D(y')\,\kappa \,y']^2 }  
\left(-1 + \sqrt{ 1 + 4 [D(y')\,\kappa \, y']^2 Re_{\tau \Delta}^2} \right),  
\label{eq:dhatudy}
\ee
where $D(y') = 1-\exp(-y' Re_{\tau\Delta}/25)$ and with a single boundary condition ${\hat u}(0)=0$. 

\begin{figure}
        \centering
        \subfloat[\label{fig:fig2}]{\includegraphics[width=0.5\columnwidth,trim=4 4 4 4,clip]{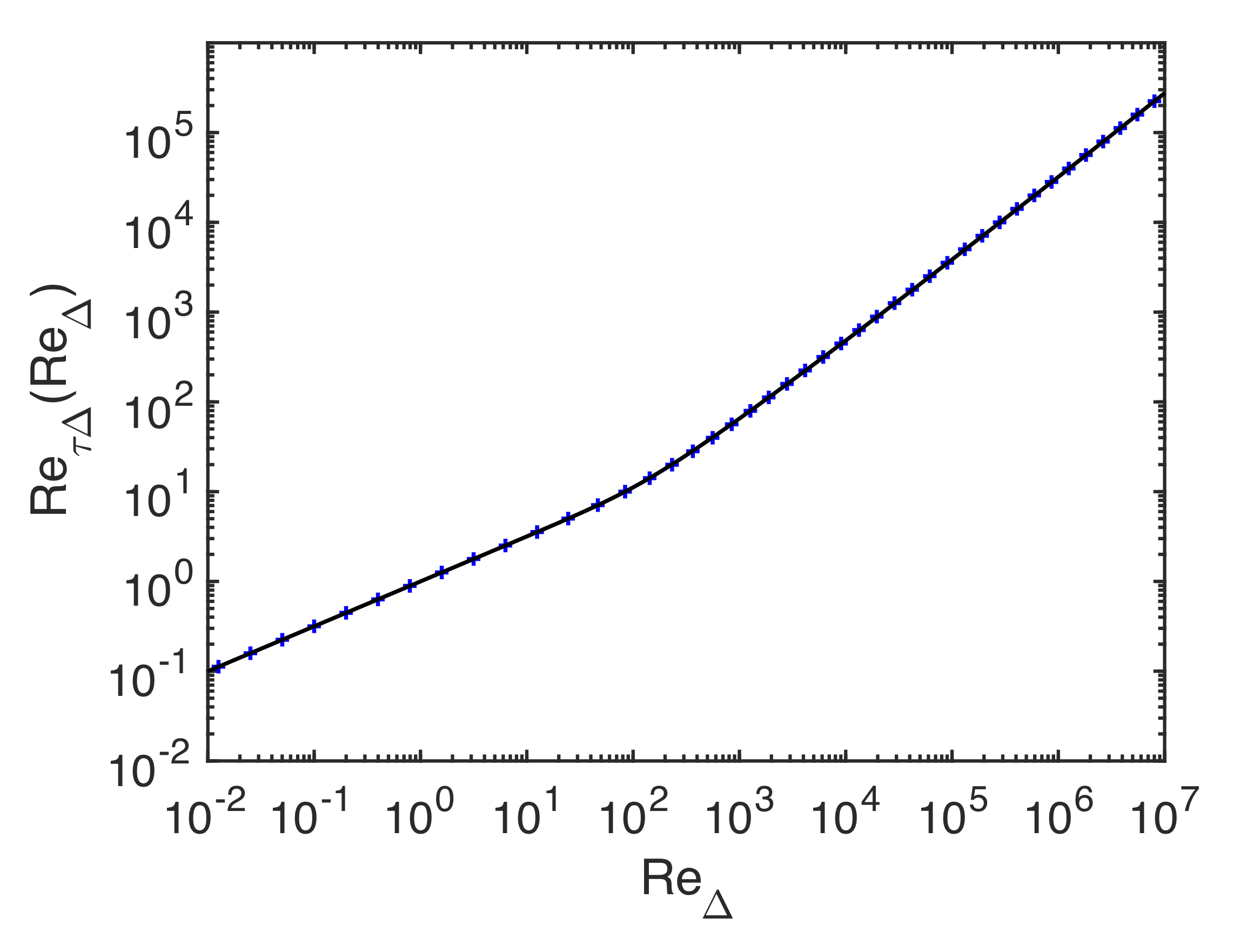}}
        \hfill
       \subfloat[\label{fig:dsd_3_5}] {\includegraphics[width=0.5\columnwidth,trim=4 4 4 4,clip]{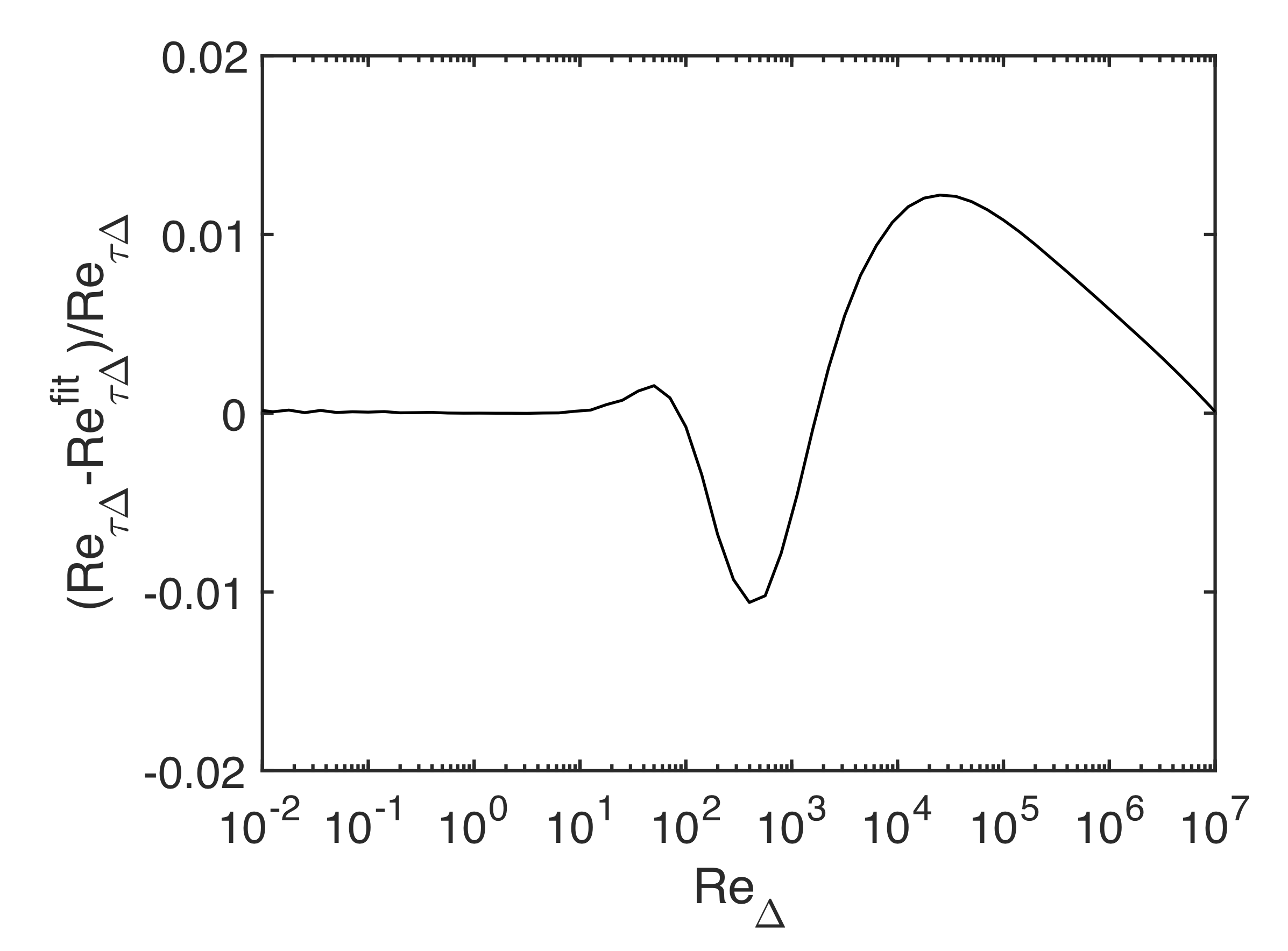}}
  \caption{ 
  (a) Blue crosses: numerical solution of Eq. \ref{eq:dhatudy} over wide range of conditions. Dark solid line: empirical fit given by Eq. \ref{eq:firstfit} with parameters given by Eqs. \ref{eq:firstparams}. (b) Relative error between numerical solution of Eq. \ref{eq:dhatudy} and empirical fit given by Eq. \ref{eq:firstfit}.
  } 
    \label{fig1}
\end{figure}

Since $D(0)=0$, we initialize at $y_i^+=10^{-3}$ or $y'_i=10^{-3}Re_{\tau\Delta}^{-1}$. The corresponding value of ${\hat u}(y'_i)$ is obtained from the near wall behavior $u(y) =  ({u_{\tau}^2 / \nu}) \, y$ or ${\hat u}(y'_i) = Re_{\tau\Delta}^2  y' _i$. The integration is done numerically (Matlab$^{\rm TM}$ ODE45), for a wide range of given $Re_{\tau\Delta}$, between $10^{-1}$ and $10^6$. The  forward integration is done until $y'=1$ is reached. The value obtained as a result, ${\hat u}(1)$, corresponds to the LES velocity normalized by $\Delta$ and $\nu$. That is to say, we find $Re_\Delta = {\hat u}(1)$ as a result of the numerical integration.  Note that this approach is equivalent to expressing the ODE in terms of $y^+$ and then integrating from $y^+=0$ up to $y^+ = Re_{\tau \Delta}$, where $Re_{\tau \Delta}$ could again be prescribed. The results of the numerical integration are shown as symbols in Fig. \ref{fig1}(a) in which $Re_\Delta$ is plotted on the x-axis and the (imposed) parameter $Re_{\tau\Delta}$ on the y-axis.  At small Reynolds numbers, the expected trend is $Re_{\tau\Delta} \sim Re_{\Delta}^{1/2}$ ($\Delta_y$ in viscous region), whereas at high $Re_{\Delta}$ the behavior is a slow approach to a linear behavior, with sub-leading logarithmic corrections (from the inverse log-law).  

Next, we aim to fit the numerical result using an empirical function. The fit function should transition smoothly between a 1/2 power law at low $Re_\Delta$ towards a power law with exponent $\beta_1$ that is on the order of 0.8-1.0 at high $Re_{\Delta}$, and which itself can be chosen to depend upon $Re_\Delta$.  We use the approach proposed by Batchelor \cite{Batchelor1951} in the context of structure function power-law transitions: 
 \be
 Re_{\tau\Delta}^{\rm fit}(Re_\Delta) \,= \,\kappa_4 \, Re_\Delta^{\beta_1} \, \left[1 + (\kappa_3 Re_\Delta)^{-\beta_2} \right]^{(\beta_1-1/2)/\beta_2}.
 \label{eq:firstfit}
 \ee
The transition sharpness is controlled by the parameter $\beta_2$.
Choosing constant values $\beta_1 = 0.9$, $\beta_2 = 1.2$, $\kappa_3 = 0.005$, and $\kappa_4 = \kappa_3^{\beta_1-1/2}$ gives results with errors of around 5\%.
Making some of the parameters dependent on  $Re_\Delta$ leads to improved accuracy. Specifically, we choose 
  \be
 \beta_1(Re_{\Delta}) = \left(1+0.155 Re_{\Delta}^{-0.03} \right)^{-1},~~~~  \beta_2(Re_{\Delta}) = 1.7-\left(1+36 Re_{\Delta}^{-0.75}\right)^{-1}.
 \label{eq:firstparams}
 \ee

The fit is displayed as solid line in Fig. \ref{fig1}(a), showing excellent agreement with the numerical solution over many decades.  The relative error is plotted in Fig. \ref{fig1}(b). The errors for  $0<Re_\Delta<10^7$ (which should easily cover all practical applications of WMLES) are below 1.2\%. In WMLES, for a given velocity $U_{\rm LES}$, one evaluates $Re_\Delta$, then applies Eq. \ref{eq:firstfit} and determines the friction velocity according to
\be
u_\tau \,=\, Re_{\tau\Delta}^{\rm fit}(Re_\Delta)  \times \frac{\nu}{\Delta_y} \,= \,U_{\rm LES} \, \frac{Re_{\tau\Delta}^{\rm fit}}{Re_\Delta}.
\ee
Thus, Eq. \ref{eq:firstfit} constitutes an equilibrium wall model for flow over smooth walls that merges with the viscous behavior and does not require iteratively solving for $u_\tau$ or numerically integrating an ODE. It does not, however, include effects of pressure gradients or roughness, considered in the next sections.
 
\section{Effects of mild pressure gradients over smooth walls}
\label{sec-mild}

The topic of modifications to the law-of-the-wall and wall functions including pressure gradient effects has received considerable attention \cite{shih2002application,skote2002direct,nickels2004inner,manhart2008near,johnstone2009resilience,tardu2010wall,duprat2011wall,coleman2015direct,gonzalez2018large,adler2020wall}. As an example, since $u_\tau$ ceases to be the only relevant velocity scale, an additive approach using another velocity profile based on a ``pressure velocity $u_p$'' \cite{shih2002application,manhart2008near,gonzalez09,duprat2011wall,luchini2018structure,adler2020wall}  has been proposed. As written in Refs.\cite{shih2002application,gonzalez09,adler2020wall} it
leads to a downward shift of the mean velocity $u^+$ in the  inertial region for favorable pressure gradient (FPG) and an upward shift for adverse pressure gradient (APG), qualitatively in agreement with the expectation that the profile becomes steeper near the wall for FPG while being less steep for APG. Similar trends are obtained when integrating the momentum equation (ODE) in the near wall region including a pressure gradient while assuming that the eddy viscosity and mixing lengths are unaffected by pressure gradient.  In this section we summarize the approach of integrating the ODE and fit the corresponding inverse wall functions for applications in which the pressure gradient is mild, far from separation and close to equilibrium conditions. We underline, however, that the physics of pressure gradient effects on boundary layers presents a number of more subtle features (some of them are recalled in Appendix A). 

\begin{figure}[h]
    \centering
    \includegraphics[width=0.70\textwidth]{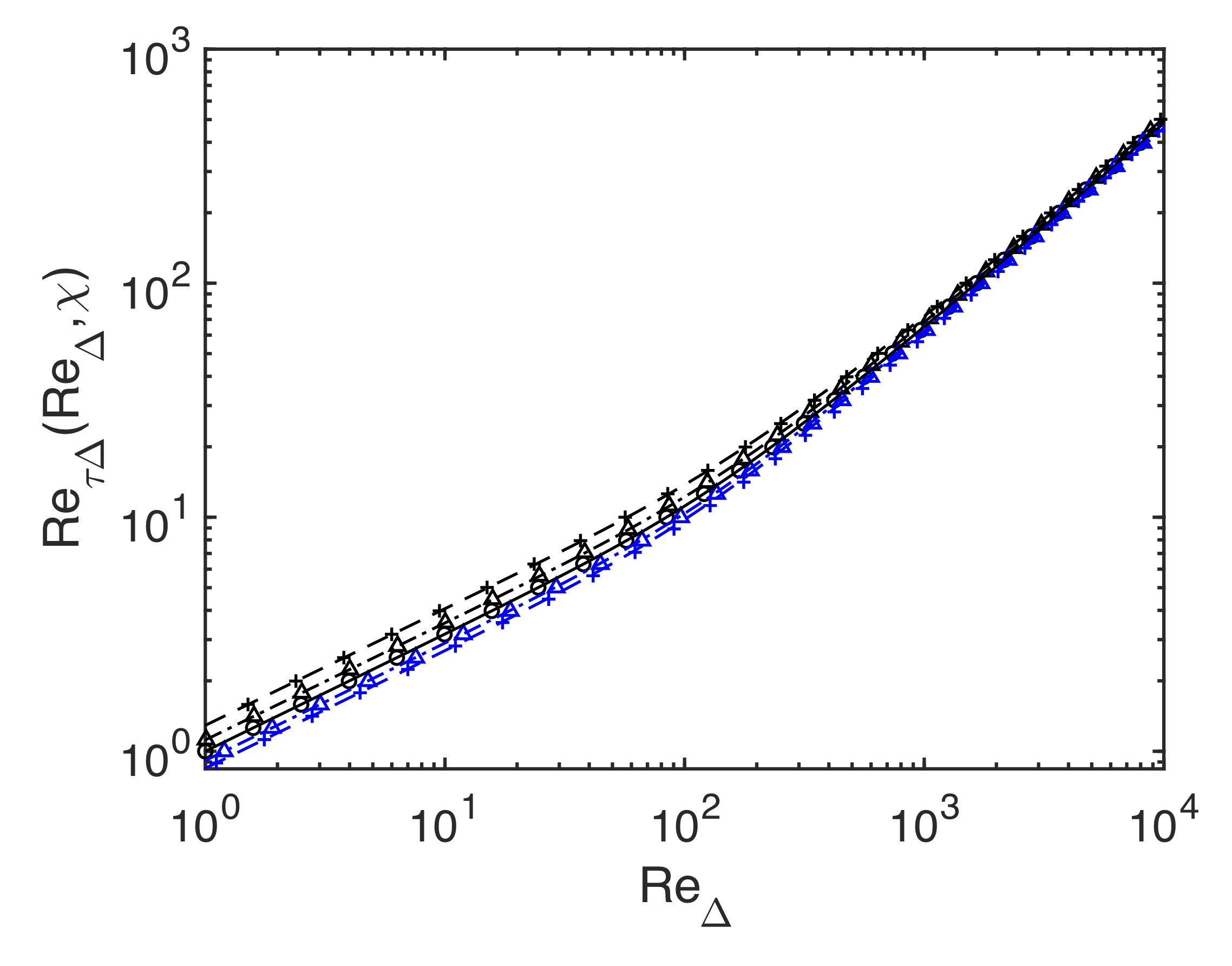}
    \caption{Symbols: numerical solution of Eq. \ref{eq:dudypress} over  a range of Reynolds numbers $Re_{\tau \Delta}$, for $\chi$ = -0.8 (black +), $\chi$ = -0.4 (black triangles), 
     $\chi$=0 (circles), $\chi$ = 0.4 (blue triangles), $\chi$ = 0.8 (blue +).  Only the region between $1<Re_\Delta<10^4$ is shown for clarity. Lines: empirical fit given by Eqs.  \ref{eq6} and \ref{eq7}. Solid line: $\chi=0$, dot-dashed lines: 
    $|\chi|=0.4$, dashed line: $|\chi|=0.8$. Black: favorable pressure gradient and ZPG cases ($\chi \leq 0$), blue lines: adverse pressure gradient $\chi>0$.}
    \label{fig2}
\end{figure}

\begin{figure}
        \centering
        \subfloat[\label{fig:fig2}]{\includegraphics[width=0.5\columnwidth,trim=4 4 4 4,clip]{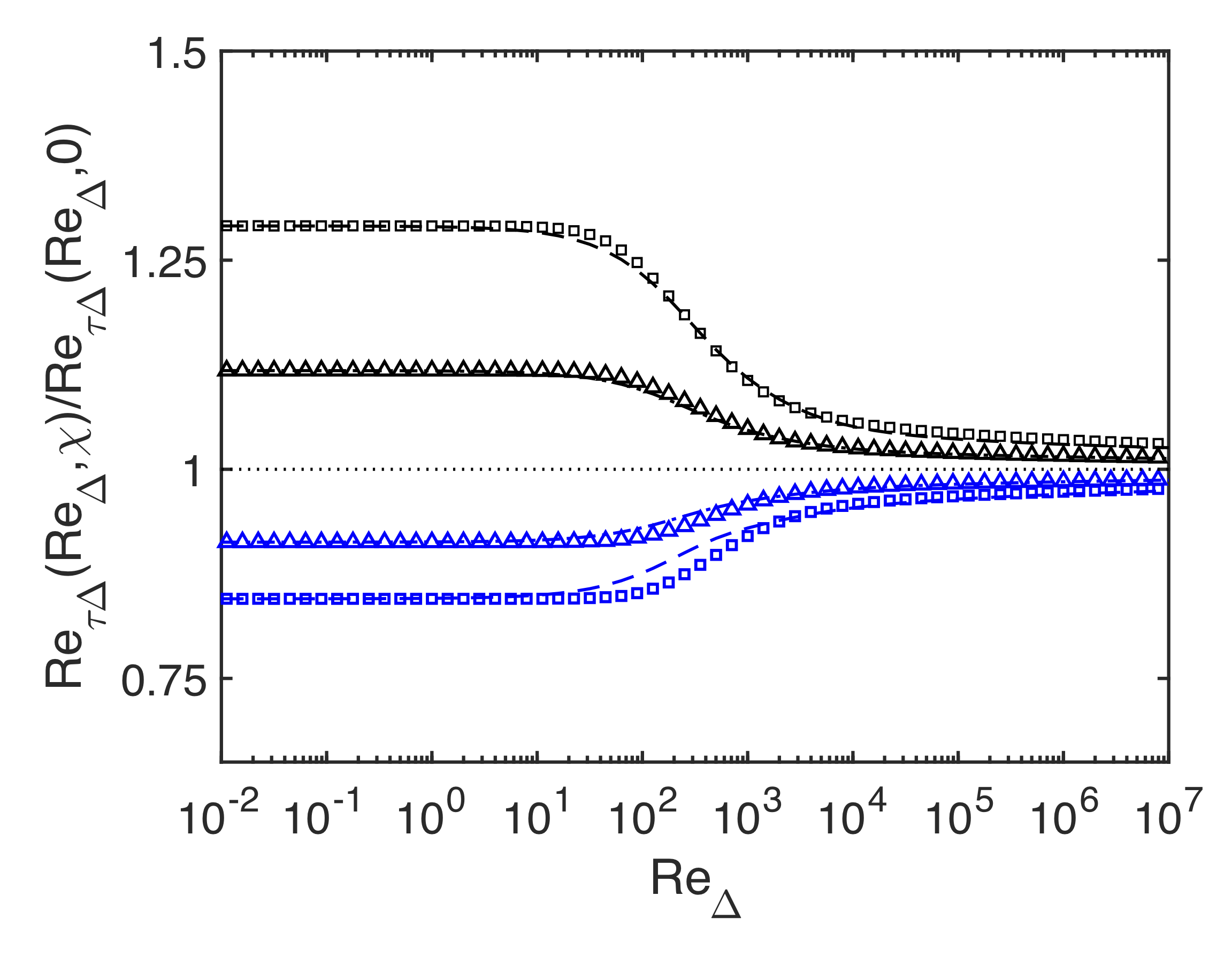}}
        \hfill
       \subfloat[\label{fig:dsd_3_5}] {\includegraphics[width=0.5\columnwidth,trim=4 4 4 4,clip]{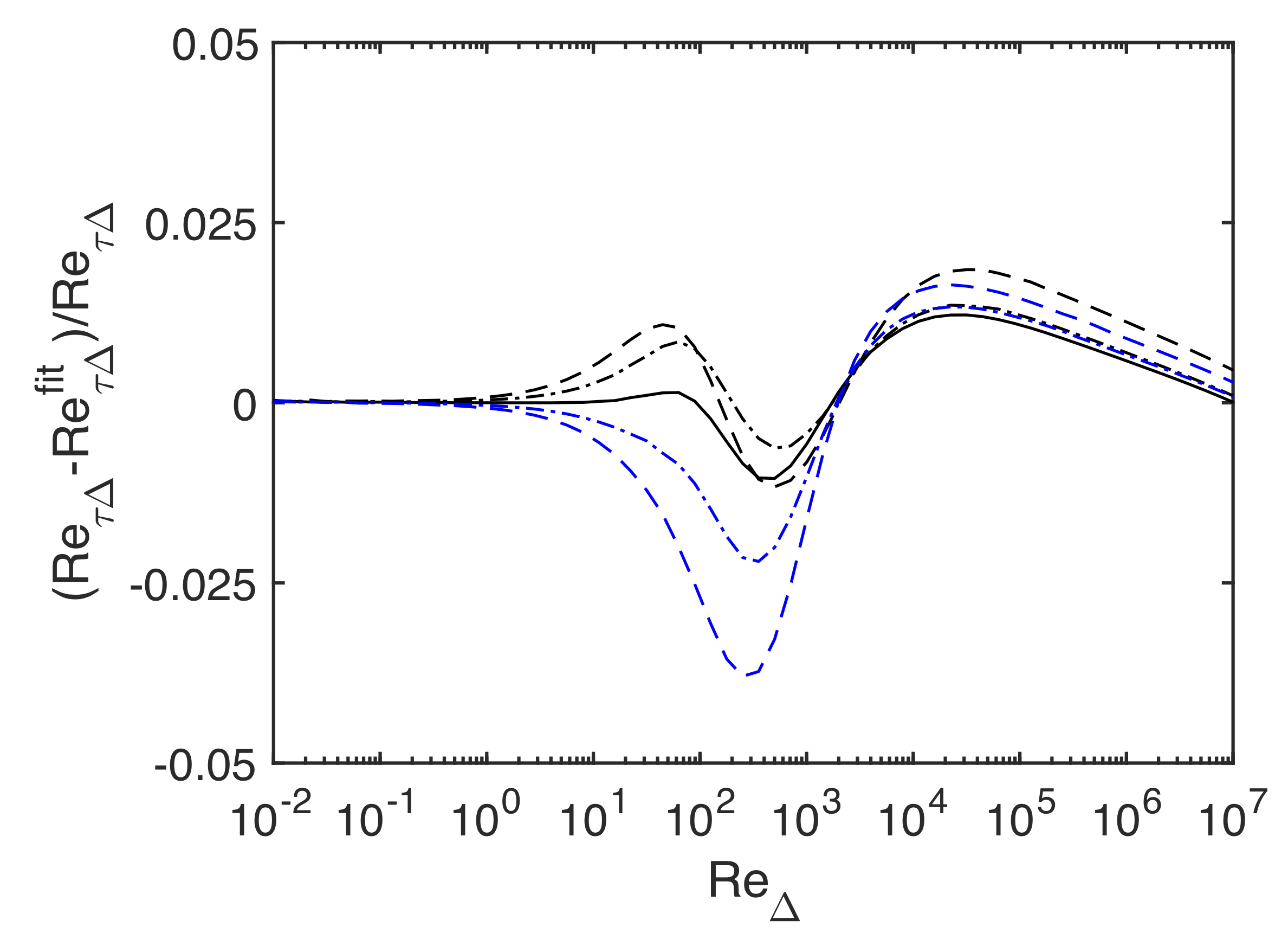}}
  \caption{ 
  (a) Symbols: Ratio of friction Reynolds number as function $Re_\Delta$ obtained from numerical integration, for:  $\chi$ = -0.8 (black squares), $\chi$ = -0.4 (black triangles), 
    $\chi$ = 0.4 (blue triangles), $\chi$ = 0.8 (blue squares). The relative effect of pressure gradient is larger at lower Reynolds number. 
    The lines are from an empirical fit (Eqs. \ref{eq6},\ref{eq7}.). (b)  Relative error between numerical solution of Eq. \ref{eq:dudypress} and empirical fit given by Eqs.  \ref{eq6} and \ref{eq7}. Solid line: $\chi=0$, dot-dashed lines: $|\chi|=0.4$, dashed line: $|\chi|=0.8$. Black: favorable pressure gradient and ZPG cases($\chi \leq 0$), blue lines: adverse pressure gradient $\chi>0$.
  } 
    \label{fig3}
\end{figure}

Defining the streamwise pressure gradient term available from LES as $N = \rho^{-1}\partial \tilde{p}_{\rm LES}/\partial s$ (for notational consistency with Ref. \cite{yang2015integral}), and again considering the momentum equation written with eddy viscosity and integrating once, yields
\be
\left(\nu+[D(y) \, \kappa \,y]^2 \left| \frac{du}{dy} \right| \right) \frac{du}{dy} = N y + u^2_{\tau}.
\ee
We neglect the effects of pressure gradient on the eddy viscosity
(see Ref. \cite{duprat2011wall} as a study where such effects are included). For a favorable pressure gradient ($N<0$), for there to be no sign changes in the slope of the velocity profile between the wall and $y=\Delta_y$, the following must hold: 
\be 
|N| < \frac{u^2_{\tau}}{\Delta_y}.
\ee
The normalized equation, after solving again the quadratic equation, reads:
\be 
\label{eq:dudypress}
\frac{d\hat u}{dy'} =  \frac{1}{2 [D(y')\,\kappa \, y']^2 }  
\left(-1 + \sqrt{ 1 + 4 [D(y')\, \kappa \,y']^2 Re_{\tau \Delta}^2  (1 + \chi \,y'\,) } \right),
\ee
where we have defined the pressure gradient parameter according to
\be
\chi = \frac{N \Delta_y}{u_{\tau}^2}
\ee
and it is understood that the developments below require $|\chi|<1$.  The boundary condition is, again, ${\hat u}(0)=0$. We initialize at  $y_i^+=10^{-3}$ or $y'_i=10^{-3}Re_{\tau\Delta}^{-1}$. The corresponding value of ${\hat u}(y'_i)$ can be obtained  from the quadratic expansion near the origin now including pressure gradient:
\be 
u(y) = \frac{u_{\tau}^2}{\nu} \, y \,+ \, \frac{N}{2 \nu} \, y^2 \, + ...
\label{visc}, ~~~~~{\rm or} ~~~~~
{\hat u}(y'_i) = Re_{\tau\Delta}^2 \left( y' _i \, + \, \frac{1}{2} \chi {y'_i}^2+ ...\right)
\ee
The integration is done  numerically as before,  obtaining $Re_\Delta = {\hat u}(1)$. The operation is repeated for a range of values of $Re_{\tau\Delta}$ and $\chi$. 
The results are shown using symbols in Fig. \ref{fig2}. 
 The effect of pressure gradient can be more readily appreciated by comparing to the $\chi=0$ case, by plotting the ratio $Re_{\tau\Delta}(Re_\Delta,\chi)/Re_{\tau\Delta}(Re_\Delta,0)$, see Fig. \ref{fig3}(a).
 
In order to ``invert'' these results we again propose an empirical fit that will now also depend on $\chi$. We note that $\chi = N \Delta / u_{\tau}^2$, and since $u_{\tau}$ is not known a-priori, $\chi$ cannot be directly evaluated in LES.  However since the effect of $\chi$ on $Re_{\tau \Delta}$ is relatively weak, in LES we may evaluate $\chi$ using 
\be 
\chi \approx \frac{N \Delta_y}{u_{\tau 0}^2}, 
\ee
where $u_{\tau 0}$ is based on $U_{\rm LES}$ only, i.e. $u_{\tau 0} = U_{\rm LES}  {Re^{\rm fit}(Re_\Delta)}/{Re_\Delta}$, using the fit of Eq. \ref{eq:firstfit} assuming $\chi=0$ as a first guess:
\be
\chi \approx \frac{N \Delta_y}{U_{\rm LES}^2} \left(\frac{Re_\Delta}{Re^{\rm fit}_{\tau\Delta}(Re_\Delta)}\right)^2.
\label{eq:chipractical}
\ee

In order to develop fits to the dependence of $Re_{\tau\Delta}$ on $Re_\Delta$ for $\chi \neq 0$, it is instructive to consider the two asymptotic limits at low and high $Re_{\Delta}$. In the viscous range (i.e. if $Re_{\tau\Delta} <<10$) 
we can obtain from Eq. \ref{visc} (using $y'=1$):
\be 
Re_{\Delta} = Re_{\tau\Delta}^2 \left( 1 + \, \frac{1}{2} \chi   \right).
\ee
But also, $Re_{\Delta} = Re_{\tau\Delta}^2(\chi=0)$, i.e. the value for $\chi=0$. We may obtain  $Re_{\tau\Delta}$ via the baseline fit in Eq. \ref{eq:firstfit} for $\chi=0$ at small $Re_{\Delta}$.
 Hence we write as the viscous limiting behavior:
 \be 
 Re_{\tau\Delta,v} = Re_{\tau\Delta}(\chi=0)  \left( 1 + \, \frac{1}{2} \chi   \right)^{-1/2}=Re_{\tau\Delta}^{\rm fit}(Re_\Delta)  \left( 1 + \, \frac{1}{2} \chi   \right)^{-1/2}.
 \ee
 
Next, we consider the limiting behavior of the solution in the inertial layer far above the viscous region, i.e. when viscosity can be neglected. The ODE simplifies to
\be 
 \frac{d\hat u}{dy'}= \frac{Re_{\tau \Delta}}{\kappa y'}  \sqrt{1+\chi\, y'\,} \,  \approx Re_{\tau \Delta}\left(\frac{1}{\kappa y'}  + \frac{1}{2\kappa}  \chi \right) .
\ee
where we have made the further assumption that $|\chi| <<1$ so that $\sqrt{1+\chi y'} \approx 1 + \frac{1}{2}  \chi y'$. Integration yields
\be 
{ \hat u}(y') =  Re_{\tau \Delta}\left(\frac{1}{\kappa} \log y'  + \frac{1}{2 \kappa}  \chi y' + C_1\right).
\ee
Consistent with the Ansatz used in the integral wall model (iWMLES) of Ref. \cite{yang2015integral}, pressure gradient effects are seen to add a linear term to the profile. Using the condition that ${\hat u}(1)=U_{\rm LES} \Delta/\nu = Re_{\Delta}$ yields 
\be
{\hat u}(y') =  Re_{\Delta}-Re_{\tau \Delta}\left(\frac{1}{\kappa} \log (1/y')  + \frac{1}{2 \kappa}  \chi (1-y') \right).
 \ee
We recall that this assumes that $\Delta_y$ is in the log-region, since molecular viscosity has been neglected.  Another condition must be invoked to determine $Re_{\tau \Delta}$ given a value of $Re_{\Delta}$. Specifically we match with the viscous behavior 
\be
{\hat u}(y') = Re_{\tau\Delta}^2  y',
 \ee
at $y^+= 11$ or $ y'  = 11/Re_{\tau\Delta}$. We note that inclusion of the pressure gradient affected second-order term and matching at the height suggested by Nickels \cite{nickels2004inner} yields only a very small correction, and will be neglected. Isolating $Re_\Delta$ and using the fact that
$11-\kappa^{-1} \log(11) =  B$ for $\kappa=0.4$ and $B=5$, we obtain
\be
\label{eq:redelta1B}
Re_{\Delta} =  Re_{\tau \Delta}\left(\frac{1}{\kappa} \log Re_{\tau\Delta}  + B + \frac{1}{2 \kappa}  \chi (1-11/Re_{\tau\Delta}) \right).
 \ee
We can also use this expression to deduce the asymptotic behavior at large $Re_{\tau\Delta}$ by using the already developed fit  $Re_{\tau\Delta}^{\rm fit}(Re_\Delta)$ as follows.
Rewrite Eq. \ref{eq:redelta1B} as 
\be
\label{eq:redelta1*}
Re^*_{\Delta} \equiv Re_{\Delta} - Re_{\tau \Delta} \frac{1}{2 \kappa}  \chi (1-11/Re_{\tau\Delta}) =  Re_{\tau \Delta}\left(\frac{1}{\kappa} \log Re_{\tau\Delta}  + B \right).
 \ee
When applied to the logarithmic layer at large $Re_{\tau\Delta}$, the fitting formula Eq. \ref{eq:firstfit} can be regarded as inverting this log-law. Thus it can now be applied in the inertial layer according to Eq. \ref{eq:redelta1*} to solve for $Re_{\tau\Delta}$ for a given $Re^*_\Delta$, i.e. to obtain $Re_{\tau\Delta,{\rm in}}=Re^{\rm fit}_{\tau\Delta}(Re^*_{\Delta})$ as function of $Re^*_\Delta$, where $Re^*_\Delta$ takes the place of $Re_\Delta$ in Eq. \ref{eq:firstfit}. Moreover, to smoothly merge to zero when $Re_{\tau\Delta}<11$ we multiply the entire additive term by a factor that tends to zero when $Re_{\tau\Delta}$ becomes smaller than O(10):
\be
Re^*_{\Delta} = Re_{\Delta} - Re^{\rm fit}_{\tau \Delta} \frac{1}{2 \kappa}  \chi 
  \left(1-\frac{11}{Re^{\rm fit}_{\tau\Delta}} \right) \left[1+\left( \frac{50}{Re^{\rm fit}_{\tau\Delta}}\right)^2 \right]^{-1/2 } .
\label{eq7}
\ee
Since the additive term depends upon the unknown  value of $Re_{\tau \Delta}$, it has been written here in terms of the fitted value for $\chi=0$, i.e. $Re^{\rm fit}_{\tau \Delta}(Re_\Delta)$ (Eq. \ref{eq:firstfit}).  Next, we combine the viscous and inertial functions $Re_{\tau\Delta,v}$ and $Re_{\tau\Delta,{\rm in}}$ using a weighting function  $\theta(Re_{\Delta}) = (1+0.0025 Re_{\Delta})^{-1}$
according to
\be
Re^{\rm com}_{\tau\Delta}(Re_{\Delta},\chi)  =  \theta(Re_\Delta) Re_{\tau\Delta}^{\rm fit}(Re_\Delta)   ( 1 + \,   \chi/2  )^{-1/2} \, +
 \, [1-\theta(Re_\Delta) ] \, Re_{\tau\Delta}^{\rm fit}(Re^*_\Delta).
\label{eq6}
\ee
 
The lines in Figs. \ref{fig2} and \ref{fig3}(a) show the results from using $Re^{\rm com}_{\tau\Delta}(Re_{\Delta},\chi)$ to predict the friction Reynolds number ratio compared to the case with zero pressure gradient. The relative error is shown in  Figure \ref{fig3}(b), for various values of $\chi$ compared to the results from the full numerical integration of the ODE. As can be seen, errors of no more than 2.5 \% are incurred. For $|\chi|<0.4$ the errors are below 1.5\%. 

It must be stressed that these developments and fits are only valid for small $\chi$, $|\chi| <<1$. For strong pressure gradient cases, the assumption of a monotonic velocity profile below $y=\Delta_y$ and a pressure-gradient independent eddy-viscosity \cite{duprat2011wall} begin to fail. The more general case including strong pressure gradients is considered in \S \ref{sec:strongpress}. 

\section{Effects of roughness with mild pressure gradients at very high $Re_\Delta$}

The `infinite Reynolds number limit' of rough wall equilibrium wall modeling   based on the profile
$u(y) =  ({u_\tau }/{\kappa}) \log\left(y/{z_0}\right)$, evaluated at $y=\Delta_y$, can be rewritten as the `infinite' Reynolds number rough wall model: 
\be 
\label{eq:roughRe0}
Re^{\infty}_{\tau \Delta} = Re_{\Delta} \,  \frac{\kappa}{\log(\Delta_y/z_0)} .
\ee
In the high Reynolds number limit, the friction and LES velocities are linearly related (the stress is quadratic with $U_{\rm LES}$), and in terms of the limiting behavior of the fits in Eq. \ref{eq:firstfit} this would correspond to $\beta_1 \to 1$. Note that expression \ref{eq:roughRe0} is applicable only for $\Delta_y >> z_0$ and that it does not include pressure gradient effects. 

Inclusion of pressure gradient can be done simply if one assumes  that the eddy viscosity scaling is unaffected by pressure gradient in the case of rough walls. While one should keep in mind the possible pitfalls of such an assumption (see e.g. Appendix A), we proceed anyhow given a lack of well-accepted wall models for pressure gradient effects including roughness effects at high Reynolds numbers. The square root of the momentum equation with the usual mixing length model neglecting viscous effects reads 
\be
\label{eq:dudypressapprox}
(\kappa y) \, \frac{du}{dy} = u_\tau \left(1+\chi  \,\frac{y}{\Delta_y}\right)^{1/2} \approx u_\tau \left(1+\frac{\chi}{2} \, \frac{y}{\Delta_y} \right),
\ee
where the last step assumes  $|\chi|<<1$. Integrating and imposing $u(y=\Delta_y)=U_{\rm LES}$ yields:
\be
u(y) = U_{\rm LES} - u_\tau \left[ \frac{1}{\kappa} \log\left(\frac{\Delta_y}{y}\right) + \frac{\chi}{2\kappa} \left(1-\frac{y}{\Delta_y} \right) \right] .
\ee
The definition of $z_0$ is that $u(z_0)=0$, and assuming the same $z_0$ is not affected by pressure gradient, we may use this condition to solve for $u_\tau$ for a given $U_{\rm LES}$ and $z_0/\Delta$, leading to
\be 
\label{eq:reyninf}
Re^{\infty}_{\tau \Delta}(Re_\Delta,\chi,\Delta/z_0) \, = Re_\Delta \,  \left[ \frac{1}{\kappa} \log\left(\frac{\Delta_y}{z_0}\right) + \frac{\chi}{2\kappa} \left(1-\frac{z_0}{\Delta_y} \right) \right]^{-1}.
\ee

As before, to evaluate $\chi$ we can use the baseline friction velocity neglecting pressure gradient, i.e. 
\be
u_{\tau 0} = U_{\rm LES} \left(\frac{\kappa}{\log(\Delta_y/z_0)} \right).
\ee
for the rough wall case (Eq. \ref{eq:roughRe0}). In general, to merge with the smooth wall behavior, one would pick the larger of the two friction velocity estimates, so that we now define the $\chi$ parameter as 
\be 
\label{eq:chipractical}
\chi = \frac{N \Delta_y} {u_{\tau 0}^2}, \,\,\,\,\,\, {\rm where}   \,\,\,\,\, u_{\tau 0} = U_{\rm LES} \, \max\left[
\frac{Re_{\tau \Delta}^{\rm fit}}{Re_\Delta} \, , \,\, \frac{\kappa}{\log(\Delta_y/z_0)} \right].
\ee
As a reminder, the modeling validity is limited to $|\chi|<<1$. 
 
Finally, we combine the smooth and rough surface behaviors into a universal fit function with a fairly sharp transition as follows:
\be
 Re^{\rm uf}_{\tau \Delta} (Re_\Delta,\chi,z_0/\Delta)  \,= \left[Re^{\infty}_{\tau \Delta}(Re_\Delta,\chi_{x},z_0/\Delta)^6+Re^{\rm com}_{\tau\Delta}(Re_{\Delta},\chi)^6\right]^{1/6},
 \label{eq:univfit}
 \ee
 where $Re^{\rm com}_{\tau\Delta}$ is given by Eq. \ref{eq6} and $Re^{\infty}_{\tau \Delta}$ by Eq. \ref{eq:reyninf}.
 
 \begin{figure}
        \centering
        \subfloat[\label{fig:fig2}]{\includegraphics[width=0.5\columnwidth,trim=4 4 4 4,clip]{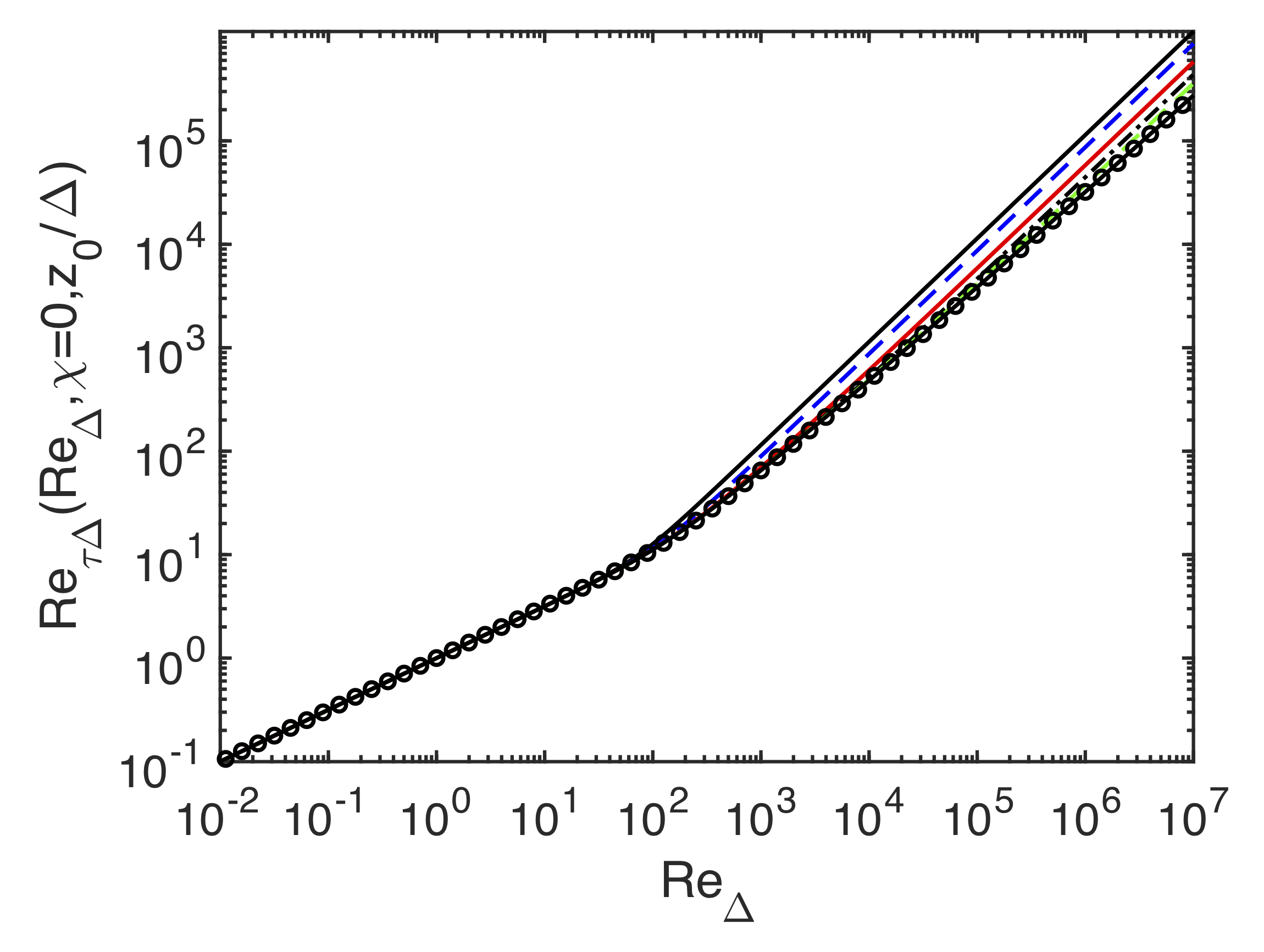}}
        \hfill
        \subfloat[\label{fig:dsd_3_5}] {\includegraphics[width=0.5\columnwidth,trim=4 4 4 4,clip]{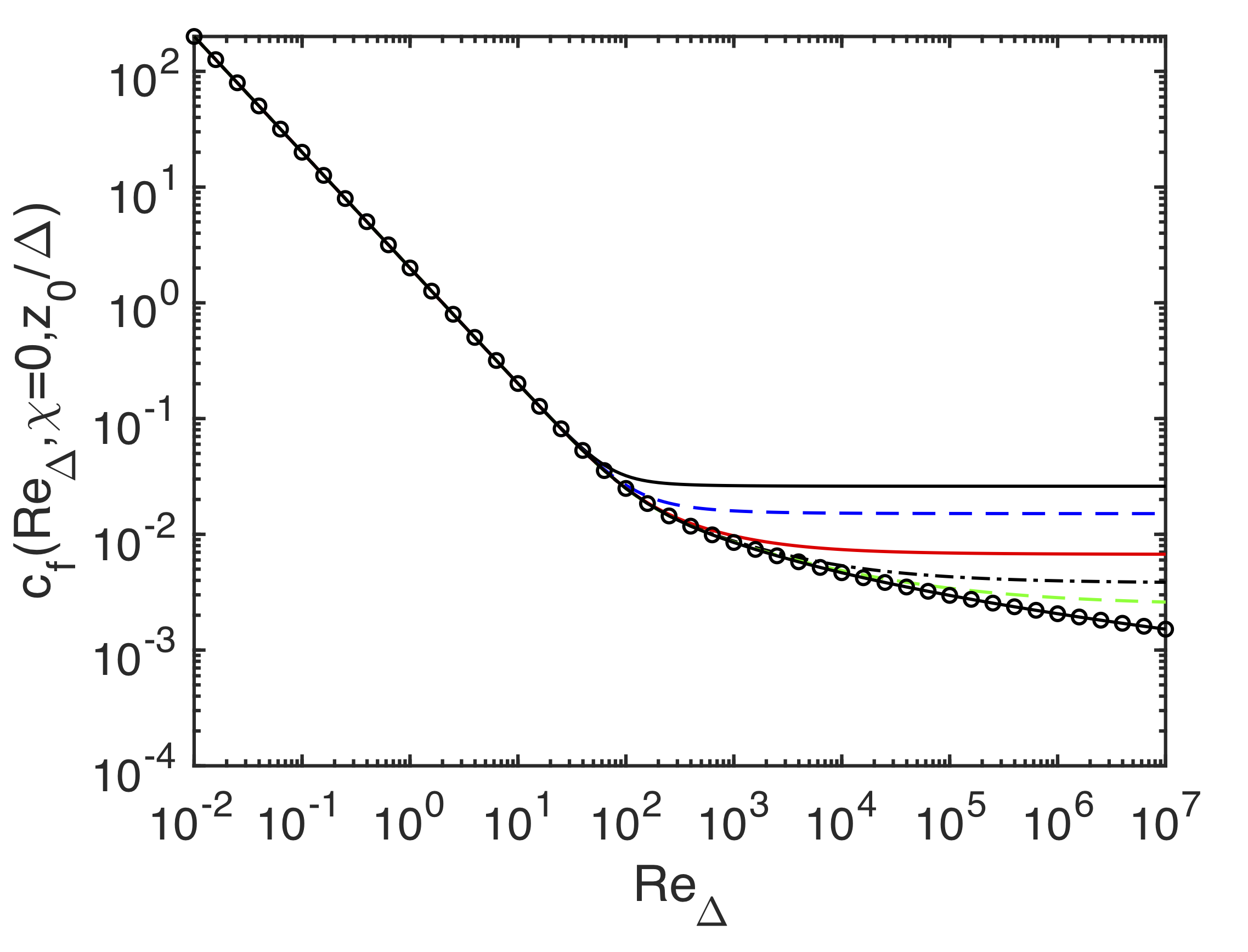}}
  \caption{ 
  (a) Lines: empirical fit from the universal fitting function  Eq.  \ref{eq:univfit} for $\chi=0$ for various values of roughness: $z_0/\Delta=3\times 10^{-2}$ (black line),  $z_0/\Delta=10^{-2}$ (blue dashed line),  $z_0/\Delta=10^{-3}$  (red line),  $z_0/\Delta=10^{-4}$ (black dot dashed line),  $z_0/\Delta=3\times 10^{-5}$ (green dashed line), smooth surface case with $z_0/\Delta \to 0$ (circles and black line). (b) Wall model Moody diagram: Friction factor from universal fitting function for $\chi=0$ for various values of roughness: $z_0/\Delta=3\times 10^{-2}$ (black line),  $z_0/\Delta=10^{-2}$ (blue dashed line),  $z_0/\Delta=10^{-3}$  (red line),  $z_0/\Delta=10^{-4}$ (black dot dashed line),  $z_0/\Delta=3\times 10^{-5}$ (green dashed line), smooth surface case with $z_0/\Delta \to 0$ (circles and black line).
  } 
    \label{fig4}
\end{figure}
 
 Figure \ref{fig4}(a) shows the results for $\chi=0$ for various values of the roughness parameter $z_0/\Delta$. 
 Figure \ref{fig4}(b) shows the same result expressed in terms of the more familiar friction factor 
 \be 
 c^{\rm wm}_{\rm f} =\frac{u_{\tau}^2}{ \frac{1}{2} U_{\rm LES}^2} = 2 \left(\frac{Re_{\tau\Delta}}{Re_\Delta} \right)^2,
 \ee
 resulting in a `generalized wall model Moody diagram'.
  
\begin{figure}
        \centering
        \subfloat[\label{fig:fig2}]{\includegraphics[width=0.5\columnwidth,trim=4 4 4 4,clip]{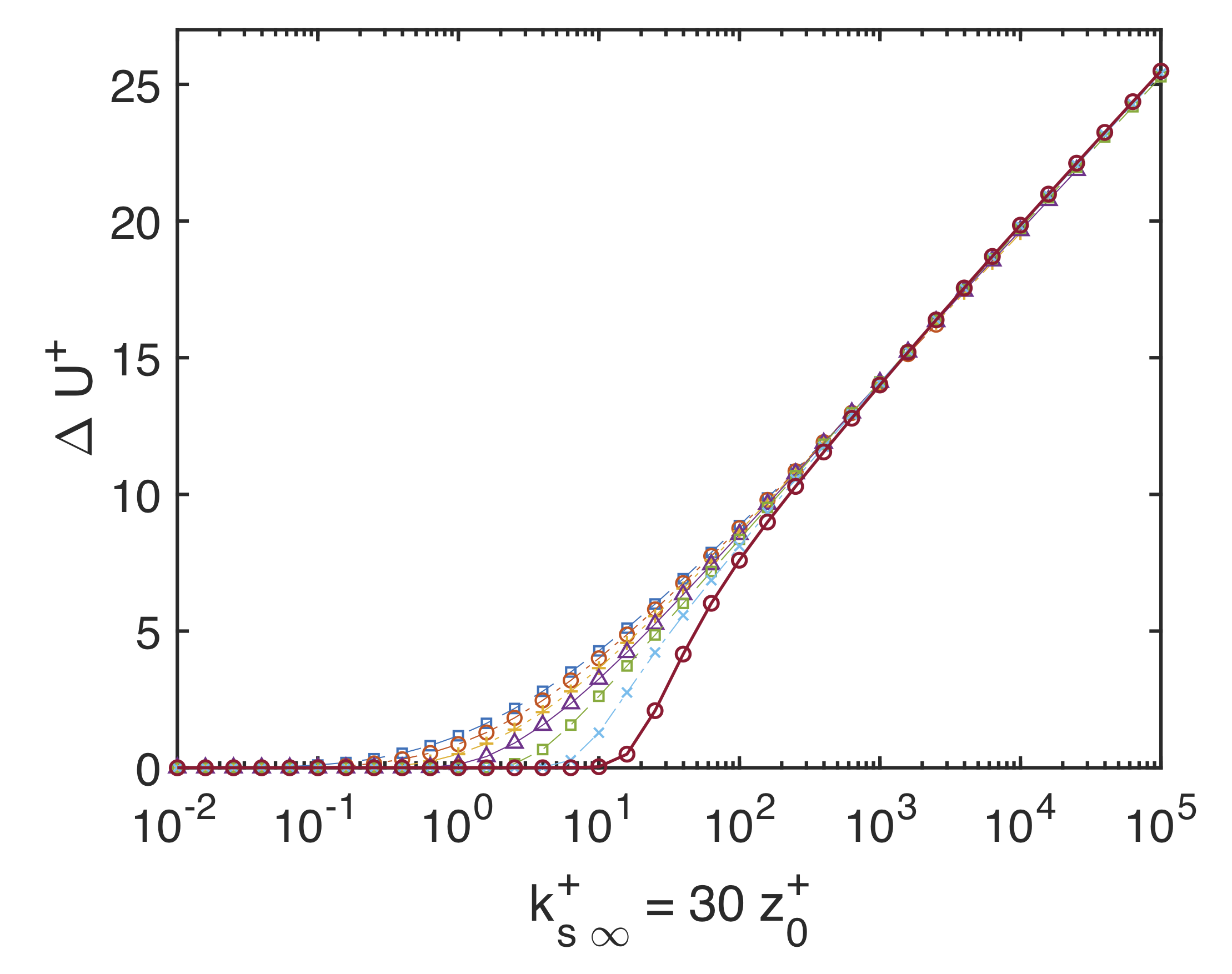}}
        \hfill
       \subfloat[\label{fig:dsd_3_5}] {\includegraphics[width=0.5\columnwidth,trim=4 4 4 4,clip]{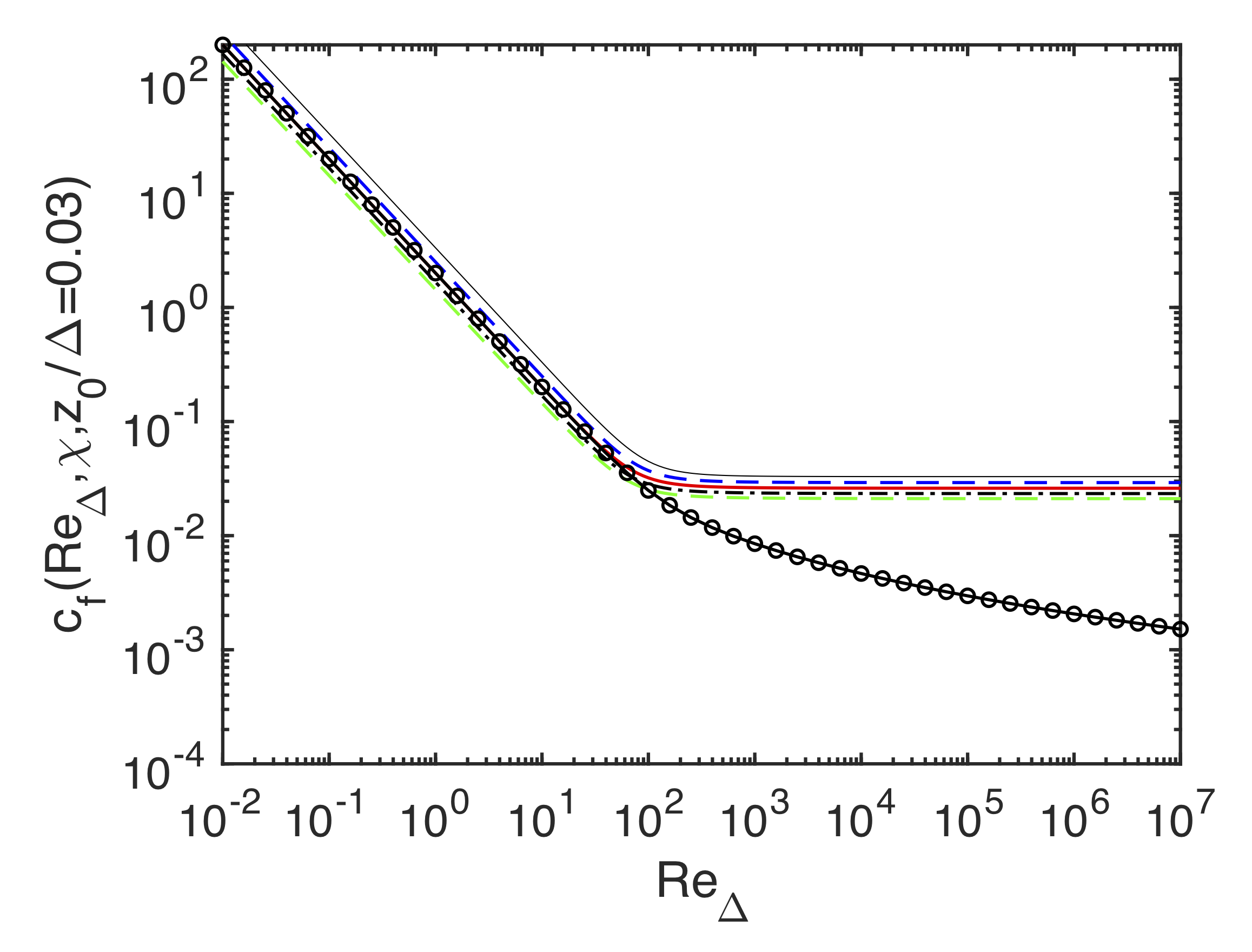}}
  \caption{ 
  (a) Roughness function $\Delta U^+$, as function of sand-grain roughness in viscous units $k_{s,\infty}^+$, for $\chi=0$ for various values of roughness: $z_0/\Delta= 10^{-1}$ (red circles and line),  $z_0/\Delta=3\times10^{-2}$ (blue crosses and dot-dashed line),  $z_0/\Delta=  10^{-2}$  (green squares and dashed line),  $z_0/\Delta=3 \times 10^{-3}$ (purple triangles and line),  $z_0/\Delta=3\times 10^{-3}$ (orange pluses and dot-dashed line), $z_0/\Delta=3 \times 10^{-4}$ (red circles and dot-dashed line)
    and $z_0/\Delta=10^{-4}$ (purple squares and dashed line).  Only values for which the resulting $Re_{\Delta,s}$ for smooth surfaces is $Re_{\Delta,s}<10^{7}$ (limits of fit) are shown. (b) Wall model Moody diagram for single roughness value at various pressure gradients. Universal fitting function for $Re_{\tau\Delta}$ as function of $Re_\Delta$ for $z_0/\Delta=3 \times 10^{-2}$ for various values of $\chi$: $\chi=-0.8$ (black line),  $\chi=-0.4$ (blue dashed line),  $\chi=0$  (red line),  $\chi=0.4$ (black dot dashed line),  and $\chi=0.8$ (green dashed line), smooth surface   ( $z_0/\Delta \to 0$) with
    $\chi=0$ (circles and black line).
  } 
    \label{fig5}
\end{figure}

Another way to display the behavior of the rough-wall fit is to compute the corresponding velocity defect, 
\be
\Delta U^+ = \frac{U_{\rm s} - U_{\rm r} } {u_\tau} =  \frac{Re_{\Delta,s}} {Re_{\tau \Delta}} - \frac{Re_{\Delta,r}} {Re_{\tau  \Delta}} 
\ee
where for a given value of $u_\tau$, $U_{\rm r}$ is the velocity at $y=\Delta_y$ corresponding to a rough surface and $U_{\rm s}$ for a smooth surface.
The sand-grain roughness in viscous units is given by the equivalency \cite{jimenez2004turbulent}, valid in the fully rough regime:
\be 
U^+ = \frac{1}{\kappa} \log\frac{y^+}{z^+_0} = \frac{1}{\kappa}  \log \frac{y^+}{k_{s,\infty}^+}  + 8.5 
\ee
which implies that
\be
k_{s,\infty}^+ = z^+_0 \exp(\kappa \,8.5) \approx 30 \,  z^+_0 \, = 30 \, \frac{z_0}{\Delta} \, Re_{\tau\Delta}.
\ee
To find $\Delta U^+$, for a given $k_{s,\infty}^+$ and ${z_0}/{\Delta}$, we first determine $Re_{\tau\Delta}= 0.0333 \, k_{s,\infty}^+ (\Delta/z_0)$. Then we invert the fit
in Eq. \ref{eq:univfit} (using {\it vpasolve} from Matlab$^{\rm TM}$) to find $Re_{\Delta,r}$ for the given ${z_0}/{\Delta}$. A second inversion is used to find $Re_{\Delta,s}$ by 
using the fit with $z_0/\Delta = 10^{-50}$, i.e. smooth surface. Only results for which $Re_{\Delta,s}<10^{7}$ (the upper limit of accuracy for the fit \ref{eq:univfit}) are plotted. 
Figure \ref{fig5}(a) displays the result (we only consider $\chi=0$ in this comparison). Comparing with Fig. 3 of Jimenez (2004) \cite{jimenez2004turbulent} (see also recent data in Ref. \cite{flack2016skin}), it can be seen that the fitting function provides realistic  predictions not only of the asymptotic behaviors at large and small $k_{s,\infty}^+$, but also for the fact that the transition between them not only depends uniquely on $k_{s,\infty}^+$ but also on additional parameters (in this case $z_0/\Delta_y$ where the transition becomes smoother for small $z_0/\Delta$ while it can be quite abrupt for larger  $z_0/\Delta$). 
  
The effects of mild pressure gradient are significant even at high Reynolds numbers for the rough surface cases. In Fig. \ref{fig5}(b) the results are shown for $z_0/\Delta = 3 \times 10^{-2}$ at various values of the pressure gradient parameter $\chi$.

\section{General pressure gradients, and flow separation on smooth surfaces}
\label{sec:strongpress}
\begin{figure}[h]
    \centering
    \includegraphics[width=0.70\textwidth]{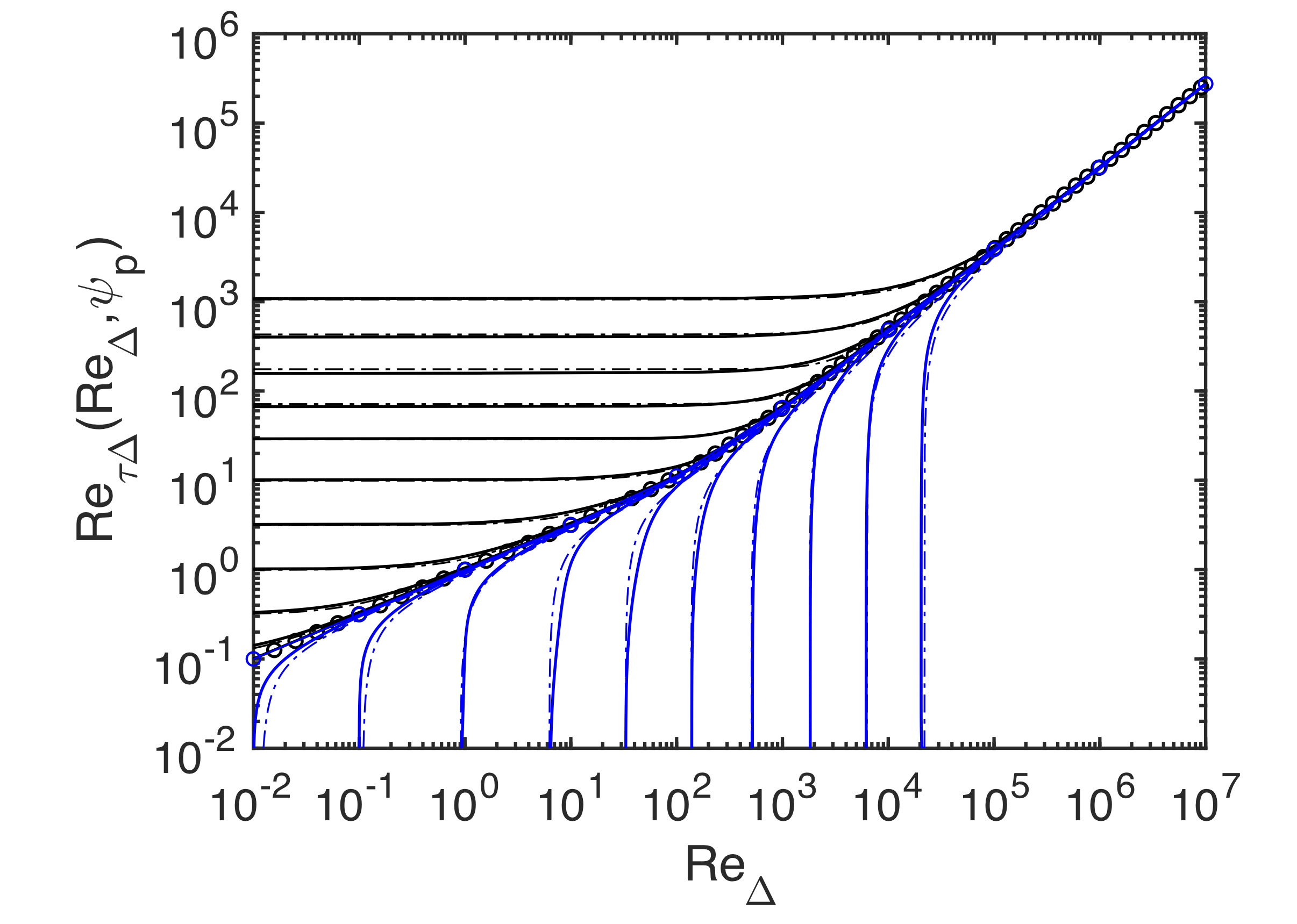}
    \caption{Solid lines: numerical solution of Eq. \ref{eq:dudypresspsi} over a range of Reynolds numbers $Re_{\tau \Delta}$, for negative pressure gradients (black lines) and positive ones (blue lines). The baseline $\psi_p=0$ case is shown with circles. The values are (from top to bottom): $\psi_p = -2 \times 10^7, -2 \times 10^6,-2 \times 10^5,-2 \times 10^4,-2 \times 10^3,-2 \times 10^2,-2 \times 10^1,-2,-0.2, -0.02, 0, 0.02, 0.2, 2, 20, 2 \times 10^2, 2\times 10^3,2 \times 10^4, 2 \times 10^5, 2 \times 10^6, 2 \times 10^7$. Dot-dashed lines: empirical fit given by Eqs.  \ref{eq:fitretaupresspsineg} and \ref{eq:fitretaupresspsipos}.}
    \label{fig:retaupresspsi}
\end{figure}

The prior sections considered the limit of small $\chi$. When the friction velocity decreases, such as approaching separation in adverse pressure gradient cases ($N>0$) or for strong favorable pressure gradient $-N\Delta_y >> u_\tau^2$, the magnitude of the parameter $\chi$ can easily exceed unity and the preceding derivations and approximations loose validity. Moreover, when $N$ becomes more dynamically relevant, scaling it with friction velocity which is itself an unknown in WMLES becomes more problematic. It is then necessary to non-dimensionalize the pressure gradient using the truly independent parameters, $\Delta_y$ and $\nu$. We thus define $\psi_p = N \Delta^3/\nu^2$ for finite Reynolds numbers (the very high Reynolds number limit is treated in the next section).  The parameter $\psi_p$ is related to $\chi$ according to $\chi = \psi_p / Re_{\tau\Delta}^2$ and with the other common pressure gradient parameter,  $p_x^+ = N \nu / u_\tau^3$, according to $p_x^+ = \psi_p / Re_{\tau\Delta}^3$. In order to determine the relationship between velocity and wall stress including strong pressure gradients, we use Eq. \ref{eq:dudypress} but rewritten according to
\be 
\label{eq:dudypresspsi}
\frac{d\hat u}{dy'} =  \frac{1}{2 [D_c(y')\,\kappa \,y']^2 }  
\left(- s + \sqrt{ 1 + 4 \, s \,[D_c(y')\,\kappa \, y']^2 (Re_{\tau \Delta}^2 + \psi_p y') } \right),
\ee
where, for $\psi_p<0$, the possibility exists that $Re_{\tau\Delta}^2/(-\psi_p)<1$ leading to a change in sign of the profile slope. For consistency with the absolute value of eddy-viscosity, for $\psi_p<0$ one must choose $s=+1$ for $0<y'\leq -Re_{\tau\Delta}^2/\psi_p$ and $s=-1$ for $-Re_{\tau\Delta}^2/\psi_p<y'\leq 1$. For $\psi_p>0$, the choice is $s=+1$. Moreover, at large pressure gradients, the van Driest damping function must be corrected. Here we use the classic Cebeci correction \cite{cebeci1970behavior} as listed in the analysis of Ref. \cite {granville1989modified}:
\be
D_c(y') = 1 - \exp\left( -  {y'Re_{\tau \Delta} \, [1+11.8 \,\psi_p \,/ \, Re_{\tau\Delta}^3]_+^{1/2}} {A^{-1}} \right).
\ee 
At this point Eq. \ref{eq:dudypresspsi} is again integrated numerically for a range of values of $\psi_p$ and $Re_{\tau\Delta}$, and the results are shown in 
Fig. \ref{fig:retaupresspsi}.

The horizontal lines for $\psi_p<0$ where $Re_{\tau\Delta} \to  Re_{\tau\Delta-{\rm min}}(\psi_p)$ becomes independent of $Re_\Delta$ correspond to the regime where the wall stress is purely determined by the favorable pressure gradient, i.e. where the relevant velocity scale is $u_p$ as used in the modeling e.g. by \cite{shih2002application,adler2020wall}. Conversely, the blue vertical lines denote the approach to  flow separation: At increasing positive $\psi_p$, there is a minimum velocity $U_{\rm LES}$, and a minimum $Re_\Delta$ ($Re_{\Delta-{\rm min}}(\psi_p)$)  required for a non-zero positive wall stress. If $Re_\Delta$ falls below this critical value, the flow has separated (or has reverse direction at the wall, a case not covered here). 

In order to fit these results, we first note that one may approximate, for $-2\times 10^7 < \psi_p <0$, the limiting asymptote as:
\be
Re_{\tau\Delta-{\rm min}}(\psi_p) = 1.5 \, (-\psi_p)^{0.39} \, \left[1+\left(\frac{1000}{(-\psi_p)}\right)^2 \right]^{-0.055}.
\ee
For $0<\psi_p<2\times 10^7$ the asymptote for $Re_\Delta$ can be fitted as:
\be
Re_{\Delta-{\rm min}}(\psi_p) =  2.5 \, \psi_p^{0.54} \, \left(1+\left[\frac{30}{\psi_p}\right]^{1/2} \right)^{-0.88}.
\ee
The fitting function for $Re_{\tau \Delta}$ in the range $-2\times 10^7 < \psi_p <0$ is:
\be
Re_{\tau \Delta}^{\rm pres} = \left( (Re_{\tau\Delta-{\rm min}}(\psi_p) )^{p(\psi_p)} + (Re_{\tau\Delta}^{\rm fit})^{p(\psi_p)} \right)^{1/p(\psi_p)},    
\label{eq:fitretaupresspsineg}
\ee
where 
$p(\psi_p) = 2.5-0.6\left[1+\tanh(2\log_{10}(-\psi_p)-6)\right]$. For the range  $0<\psi_p<2\times 10^7$  {\it and} $Re_\Delta > Re_{\Delta-{\rm min}}$, the fitting function is  
\be
Re_{\tau \Delta}^{\rm pres} =  Re_{\tau\Delta}^{\rm fit} \left(1-\frac{1}{(1+\log[Re_\Delta/Re_{\Delta-{\rm min}}(\psi_p)])^{1.9} } \right),
\label{eq:fitretaupresspsipos}
\ee
while for $Re_\Delta \leq Re_{\Delta-{\rm min}}$: 
$$Re_{\tau \Delta}^{\rm pres} = 0.$$
The fits are shown with dot-dashed lines in Fig. \ref{fig:retaupresspsi}. Overall, the agreement is good except in some regions where the slopes are very large for $\psi_p > 0$. 

\section{General pressure gradients and flow separation in fully rough regime:}
\begin{figure}
        \centering
        \subfloat[\label{fig:fig2}]{\includegraphics[width=0.55\columnwidth,trim=4 4 4 4,clip]{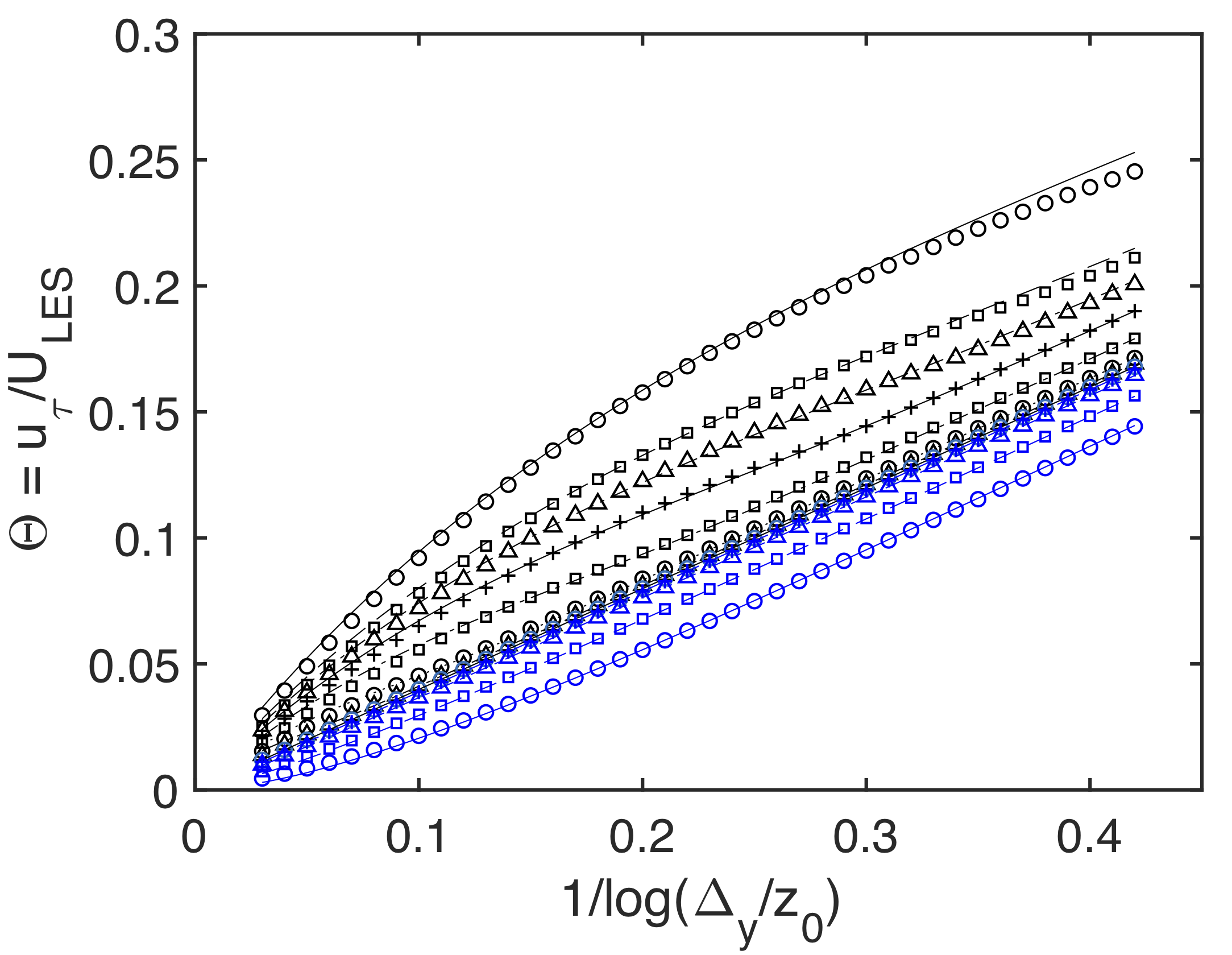}}
        \hfill
       \subfloat[\label{fig:dsd_3_5}] {\includegraphics[width=0.45\columnwidth,trim=4 4 4 4,clip]{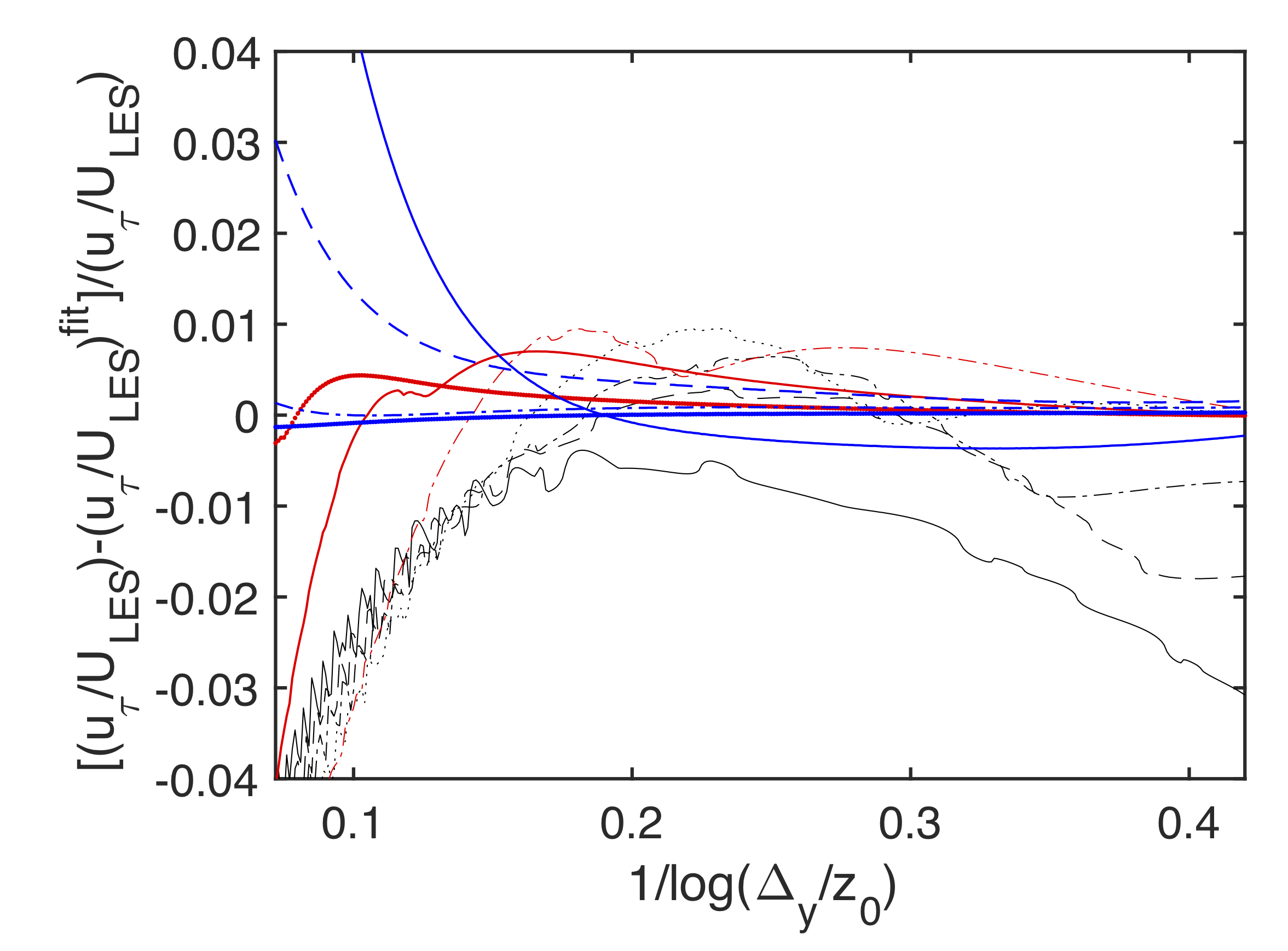}}
  \caption{ 
  (a) Symbols: numerical solution of Eq. \ref{eq:dudyroughpress} (integration and solving the condition $u'(1)=1$ for $\Theta$). Black symbols and lines: $\Psi_p<0$,  blue symbols and lines: $\Psi_p>0$. From top to bottom, the values are $\Psi_p=$ -0.07, -0.04, -0.03,   -0.02, -0.01, -0.003, -0.001, 0.0001,  0.001, 0.003, 0.01, 0.02. Lines: empirical fit given by Eq.  \ref{eq:fitutauULES}. (b)  Relative error between fit and numerical solution (symbols same as in (a)).
  } 
    \label{fig:utauULESvs1log}
\end{figure}

Lastly, we wish to include the case of high Reynolds number fully rough surfaces including possibly strong pressure gradients. Here we again use Eq. \ref{eq:dudypressapprox} but without assuming that $|\chi| <<1$ to allow for strong pressure gradients. However, unlike the situations treated earlier, now that viscosity is no longer a relevant parameter and since $u_\tau$ does not necessarily scale linearly with $U_{\rm LES}$, it is not useful to introduce the Reynolds numbers $Re_{\Delta}$ and $Re_{\tau \Delta}$. Instead, we use a dependent variable similar to the friction coefficient, the ratio $\Theta =u_\tau/U_{\rm LES}$ (as shorthand for $\sqrt{c_{\rm f}^{\rm wm}/2}$), since the only known velocity scale available for scaling is $U_{\rm LES}$. Similarly, for the pressure gradient parameter, we have only one possible choice, namely $\Psi_p = N \Delta_y / U_{\rm LES}^2$. Again using $y'=y/\Delta_y$, we rewrite the ODE to be solved according to
\be
\label{eq:dudyroughpress}
 \frac{du'}{dy'} =  \frac{1}{\kappa y'}  \Theta \left(1+ \,\frac{\Psi_p}{\Theta^2} \, y'\right)^{1/2} \,\,\,\, {\rm for} \,\,\,\, \Psi_p \geq 0 , \,\, {\rm or} \,\,   y' < \frac{\Theta^2}{(-\Psi_p)}  \,\,\,{\rm when } \,\,\, \Psi_p<0  \,  , 
\ee
$$
\frac{du'}{dy'} =  - \frac{1}{\kappa y'}  \Theta \left(-\left[1+ \,\frac{\Psi_p}{\Theta^2} \, y'\right] \right)^{1/2} \,\,\,\, {\rm for} \,\,\,\, \Psi_p < 0 \,\, {\rm and} \,\,   y' > \frac{\Theta^2}{(-\Psi_p)}\,\,\,\,\,\,\,\,\,\,\,\,\,\,\,\,\,\,\,\,\,\,\,\,\,\,\,\,\,
$$
where now $u' = u/U_{\rm LES}$. The problem can be solved by integrating this equation between $y'=z_0/\Delta_y$ (where $u'=0$) and $y'=1$ and, for a given pressure gradient parameter $\Psi_p$, find the value of $\Theta$ for which the integral yields $u'(1) = 1$. Eq. \ref{eq:dudyroughpress} admits analytical solutions with different forms (square roots, atanh(..) and  atan(..) depending on the signs of $\Psi_p$), see for instance Ref. \cite{skote2002direct}. One can then set those solutions equal to 1 and solve numerically for $\Theta$. However, the analytical solutions rely on exact cancellations near $y' \to \Theta^2/|\Psi_p|$ that are difficult to capture accurately in a subsequent numerical solution procedure. Hence it was found simpler to integrate Eqs. \ref{eq:dudyroughpress} numerically (Matlab$^{\rm TM}$'s ODE45) and then find $\Theta$ using a bisection method. The procedure is repeated for a range of pressure and roughness parameters, $\Psi_p$ and $z_0/\Delta_y$, respectively. Since for the zero-pressure gradient standard case the solution is 
$\Theta = \kappa/\log(\Delta_y/z_0)$, we plot the numerically obtained solution $\Theta(z_0/\Delta_0,\Psi_p)$ as function of $1/\log(\Delta_y/z_0)$ in Fig. \ref{fig:utauULESvs1log}. 
 
 In developing a fit, we aim to comply with the limiting scaling that when the roughness becomes small and $z_0/\Delta$ less important in determining $u_\tau$ based on $U_{\rm LES}$, we expect $\Theta \sim |\Psi_p|^{1/2}$ since the only velocity scale left is then $(N \Delta_y)^{1/2}$. A functional form that provides a good approximation is given by
 \be
 \label{eq:fitutauULES} 
 \Theta^{\rm fit}(z_0/\Delta_y,\Psi_p) = \, \kappa \, \xi \, + \theta_1(z_0/\Delta_y,\Psi_p) \, + \, \theta_2(z_0/\Delta_y,\Psi_p),
 \ee
 where
 \be 
 \theta_1(z_0/\Delta_y,\Psi_p) = - {\rm sign}(\Psi_p)\,
 \sqrt{|\Psi_p|} \,\, \alpha_\Psi\, \left[1+\left( \frac{2.25 \, \xi}{\alpha_\Psi}     \right)^{-1.35} \right]^{-1/1.35}, \,\,\,{\rm and}
 \ee
 \be
 \theta_2(z_0/\Delta_y,\Psi_p) = \,C(\Psi_p) \, \frac{\xi}{\xi_m}\, \exp\left(\frac{1}{2}\left[1-\left(\frac{\xi}{\xi_m}\right)^2 \right] \right), 
 \ee
with 
 $\xi =  [\log(\Delta_y/z_0)]^{-1}$, $\alpha_\Psi = 1.15 \, |\Psi_p|^{1/2}$,
 and for $\Psi_p<0$: $C(\Psi_p) = 0.085 \,|\Psi_p|^{1/2}$ and $\xi_m = 0.95\, |\Psi_p|^{1/2} $, while for $\Psi_p>0$: 
 $C(\Psi_p) = -0.63\, \Psi_p^{1.24} $ and $\xi_m = 0.20$.
  
 The lines in Fig. \ref{fig:utauULESvs1log}(a) indicate the fitting function. The  relative differences between the fit and the numerical solution are shown in Fig. \ref{fig:utauULESvs1log}(b).  As can be seen, the fit is less accurate than for the fits of the viscous solutions introduced in the previous sections. In the range $10^{-5} < z_0/\Delta < 0.1$ and $|\Psi_p|<0.07$ for $\Psi_p<0$ or $|\Psi_p|<0.02$ for $\Psi_p>0$, relative errors are below 4\% for $\Theta$.

\section{Combined viscous, rough surface and pressure gradient fitting function}

\begin{figure}
        \centering
        \subfloat[\label{fig:fig2}]{\includegraphics[width=0.5\columnwidth,trim=4 4 4 4,clip]{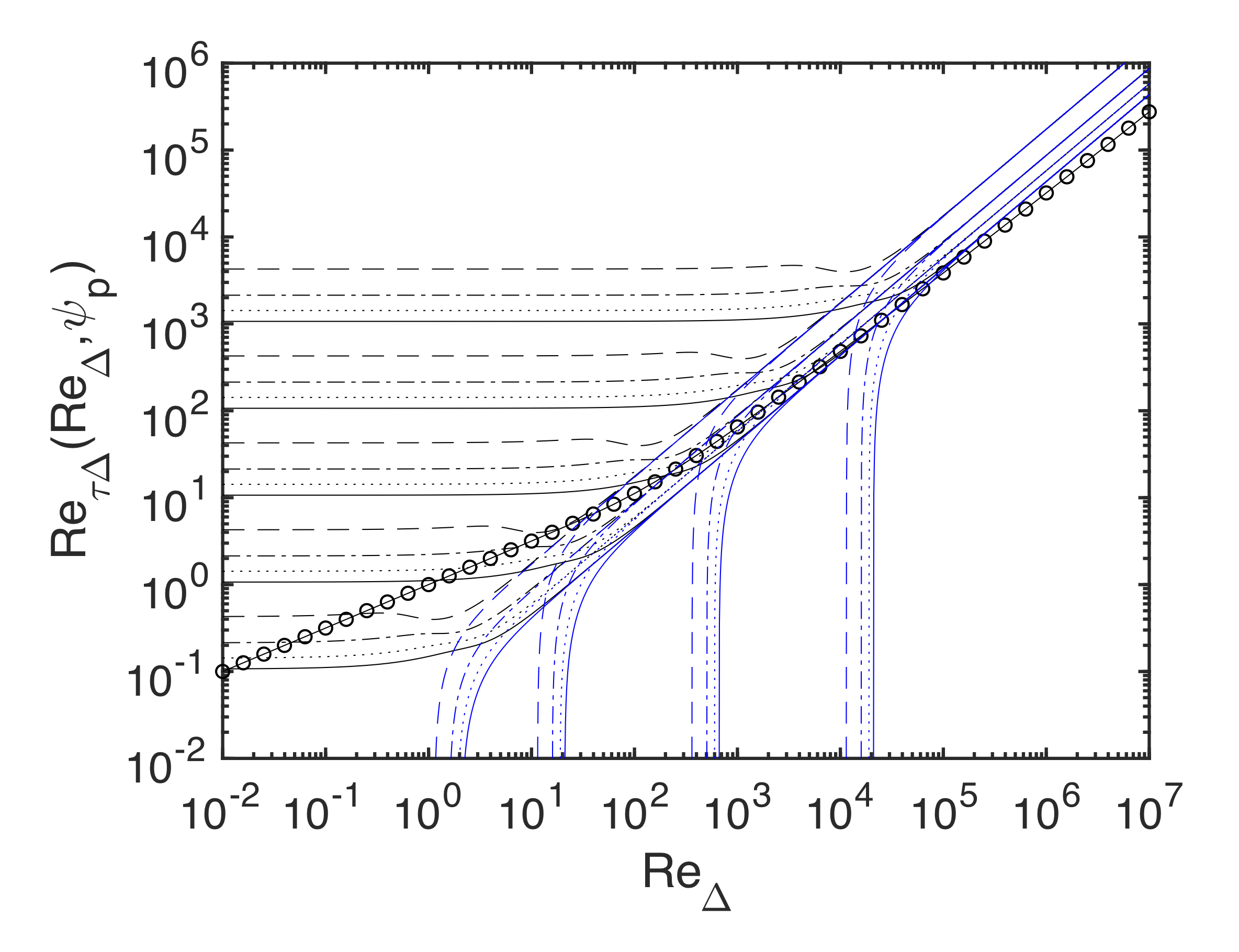}}
        \hfill
       \subfloat[\label{fig:dsd_3_5}] {\includegraphics[width=0.5\columnwidth,trim=4 4 4 4,clip]{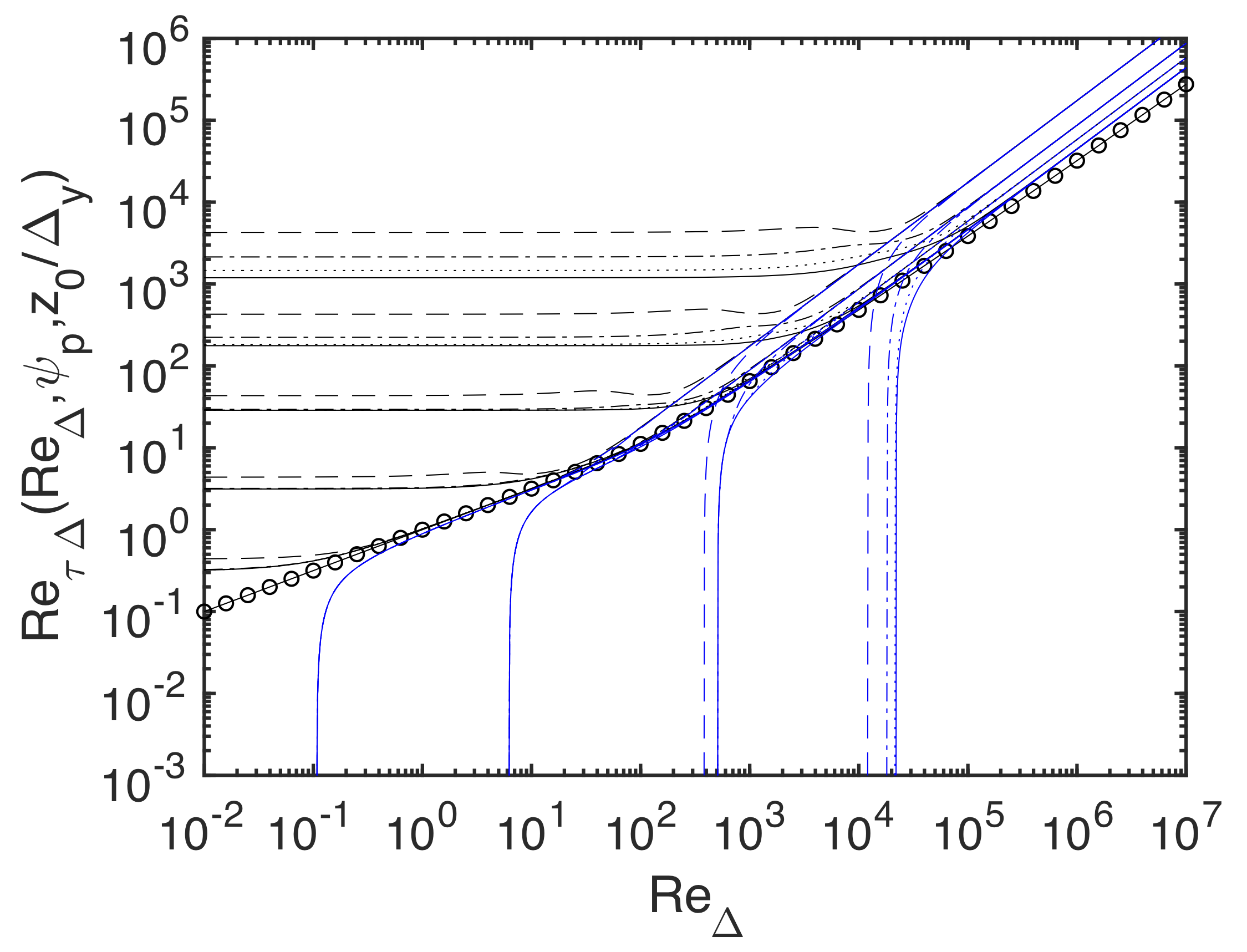}}
  \caption{ 
  (a) Fitting function (Eq. \ref{eq:fitutauULES}) for the fully rough surfaces and for strong adverse and favorable pressure gradients expressed using viscosity dependent parameters $Re_\Delta$ and $\psi_p$. $Re_{\tau} = Re_{\Delta} \Theta^{\rm fit} (z_0/\Delta_y,\Psi_p=\psi_p/Re_\Delta^2)$. 9 groups of 4 lines show results for 9 pressure gradient parameters: For favorable cases (black lines, groups from top to bottom): $\psi_p=-2\times 10^{-7}$, $-2\times 10^{-5}$, $-2\times 10^{-3}$, -20, -0.2; for adverse pressure gradients (blue lines, groups from left to right): $\psi_p = +0.20$, 20, $2\times10^4$ and  $2\times10^7$. For each group, the different line types denote different roughnesses: $z_0/\Delta_y=10^{-4}$ (solid lines),
  $z_0/\Delta_y=10^{-3}$ (dotted lines), $z_0/\Delta_y=10^{-2}$ (dot-dashed lines) and $z_0/\Delta_y=10^{-1}$ (dashed lines). As reference, open circles denote the smooth ZPG case. (b) Fitting function combining fully rough and smooth surface results, i.e. $Re_{\tau \Delta}^{\rm ufs}(Re_\Delta,\psi_p,z_0/\Delta_y)$ (lines same as in (a).  
  } 
    \label{fig:retauvsredpsiroughall}
\end{figure}

Before combining the fully rough-surface fitting function presented in the last section with the finite viscosity fits, it is of interest to first present the (infinite Reynolds number) rough-surface fit in terms of the parameters used for smooth walls to compare the results. To this effect we evaluate 
\be
Re_{\tau \Delta}(Re_\Delta,\psi_p,z_0/\Delta) \,= \,Re_{\Delta} \,\,\Theta^{\rm fit} (z_0/\Delta_y,\Psi_p=\psi_p/Re_\Delta^2). 
\ee
Results are shown in Fig. \ref{fig:retauvsredpsiroughall}(a). The expected invariance of $Re_{\tau \Delta}/Re_{\Delta}$ with constant $\Psi_p=\psi_p/Re_\Delta^2$ for any given $z_0/\Delta$ is apparent in the figure's ``translatability'' along its diagonal. The slight wiggles and non-monotonicity seen in the fits near the transition (e.g. the dashed line for $Re_{\Delta}=10^3$ and $Re_{\tau\Delta}=500$, $\psi_p=-2\times 10^{-5}$) are caused by imperfections of the proposed fitting function Eq. \ref{eq:fitutauULES} (the errors shown in Fig. \ref{fig:utauULESvs1log}b) and could be improved further although the complex dependencies with the 2 parameters make finding improved fits challenging. Training a neural net may be fruitful in this context, now that the physical trends are clearly identified.  

It is interesting to note that similar to the viscous case, for large favorable pressure gradients and low $Re_{\Delta}$, $Re_{\tau \Delta}$ becomes independent on $Re_{\Delta}$ (i.e. for a fixed $\nu$ and $\Delta_y$, the wall stress becomes independent of the velocity $U_{\rm LES}$). In this limit the velocity scale is provided by the imposed pressure gradient that drives the flow. For a fixed pressure gradient the wall stress increases with roughness, as expected. For adverse pressure gradients, the results show that there are conditions of sudden drop in wall stress, corresponding to incipient separation. That is to say, if the pressure gradient is adverse and the velocity $U_{\rm LES}$ sufficiently low the stress drops to zero. Interestingly, we see that increased roughness enables lower velocities before separation occurs, or for a fixed $Re_\Delta$ and $\psi_p$, increased roughness leads to increasing $Re_{\tau\Delta}$, consistent with the ``golf ball dimples effect''. 

Finally, we combine the finite Reynolds number smooth surface and rough wall infinite Reynolds number fits by choosing the largest of the two, with a relatively sharp transition among the two (similar to Eq. \ref{eq:univfit}):
\be
Re_{\tau \Delta}^{\rm ufs}(Re_\Delta,\psi_p,z_0/\Delta_y) = \left( Re_{\tau \Delta}^{\rm pres}(Re_\Delta,\psi_p)^6 +  
Re_{\Delta} \, \Theta^{\rm fit} (z_0/\Delta_y,\psi_p/Re_\Delta^2)^6 \right)^{1/6}.
\label{eq:ufsfit}
\ee

\begin{figure}
        \centering
        \subfloat[\label{fig:fig2}]{\includegraphics[width=0.5\columnwidth,trim=4 4 4 4,clip]{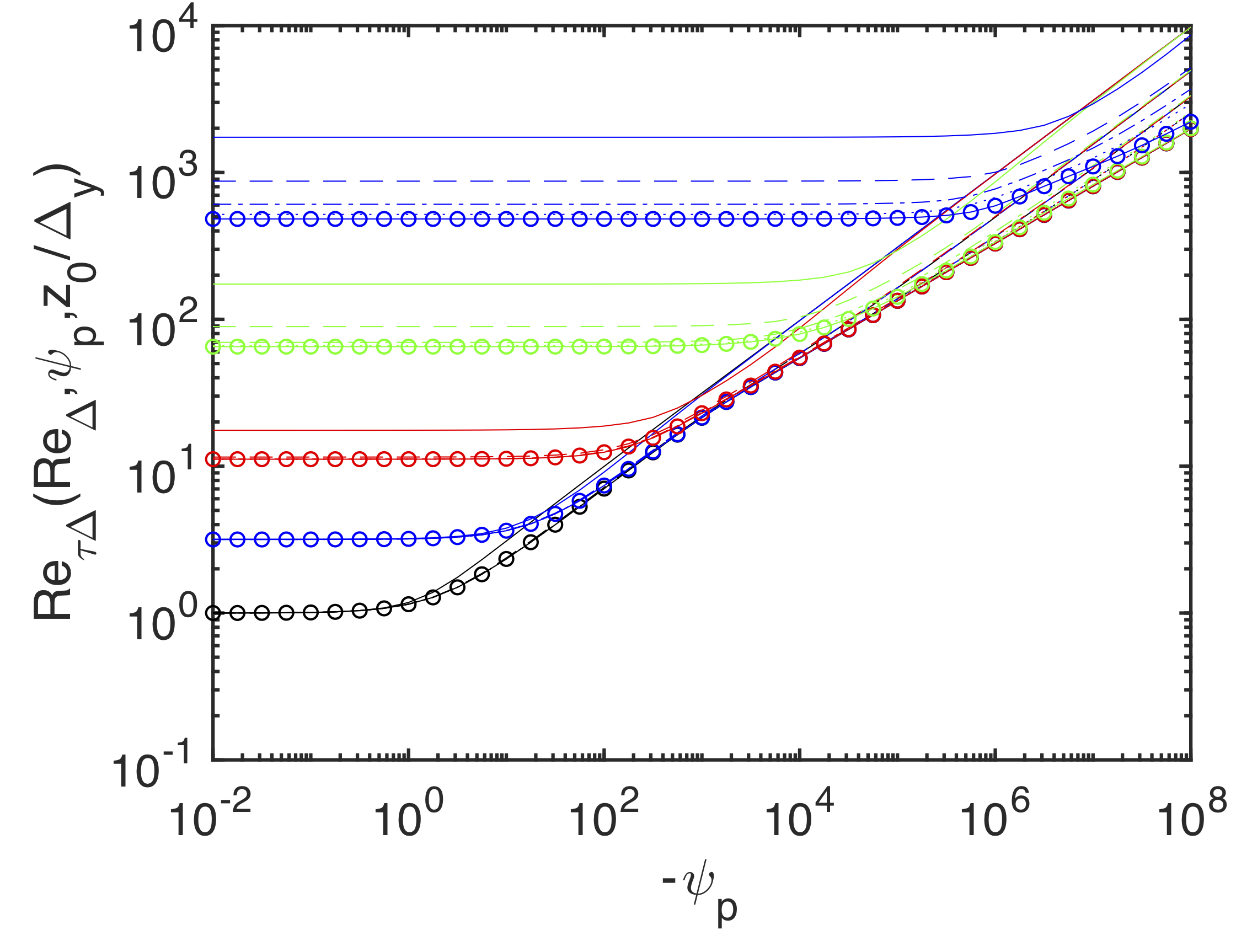}}
        \hfill
       \subfloat[\label{fig:dsd_3_5}] {\includegraphics[width=0.5\columnwidth,trim=4 4 4 4,clip]{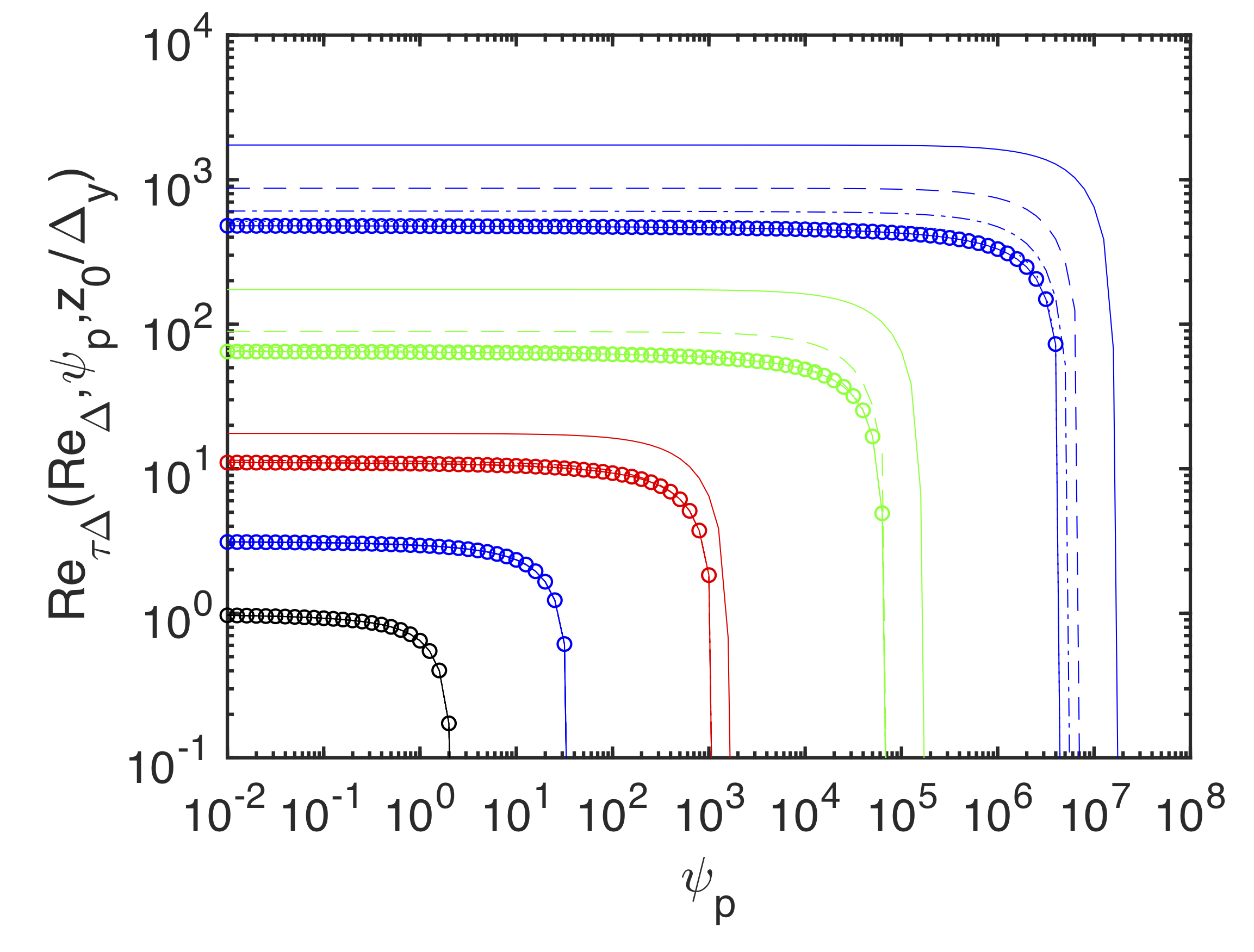}}
  \caption{(a) Fitting function combining fully rough and smooth surface results, i.e. $Re_{\tau \Delta}^{\rm ufs}(Re_\Delta,\psi_p,z_0/\Delta_y)$ as function of pressure gradient parameter $\psi_p$ for favorable pressure gradient cases ($\psi_p<0$). Five groups of curves are shown, corresponding to: $Re_\Delta = 1$ (black), $Re_\Delta = 10$ (blue),
  $Re_\Delta = 100$ (red), $Re_\Delta = 1000$ (green) and $Re_\Delta = 10000$ (blue).  For each group, the different line types denote different roughness cases: $z_0/\Delta_y=10^{-50}$ (smooth surface, solid line and circles), $z_0/\Delta_y=10^{-4}$ (dotted lines),
  $z_0/\Delta_y=10^{-3}$ (dot-dashed lines), $z_0/\Delta_y=10^{-2}$ (dashed lines) and $z_0/\Delta_y=10^{-1}$ (solid lines). (b) Fitting function $Re_{\tau \Delta}^{\rm ufs}(Re_\Delta,\psi_p,z_0/\Delta_y)$ for adverse pressure gradient $\psi_p > 0$ (line types same as in (a)).} 
    \label{fig:retauvspsi}
\end{figure}

The results in Fig. \ref{fig:retauvsredpsiroughall}(b) show how viscous effects overwhelm the lower roughness effects in the lower Reynolds number regimes. The afore-mentioned trends (constant $Re_{\tau\Delta}$ asymptote for strong favorable pressure gradient and separation for adverse pressure gradient cases) still hold, but viscous effects provide a lower bound for $Re_{\tau\Delta}$ compared to the small roughness, small $z_0/\Delta$ cases (i.e. when $z_0^+$ becomes small).  The resulting complete fitting function can also be presented as function of $\psi_p$ for various $Re_\Delta$ as shown in Fig. \ref{fig:retauvspsi} for both favorable (a) and adverse (b) pressure gradients. For large $|\psi_p|$ in the favorable pressure gradient cases, for rough surfaces the $Re_{\tau\Delta}\sim |\psi_p|^{1/2}$ scaling is apparent. 

\begin{figure}[h!]
    \centering
    \includegraphics[width=0.70\textwidth]{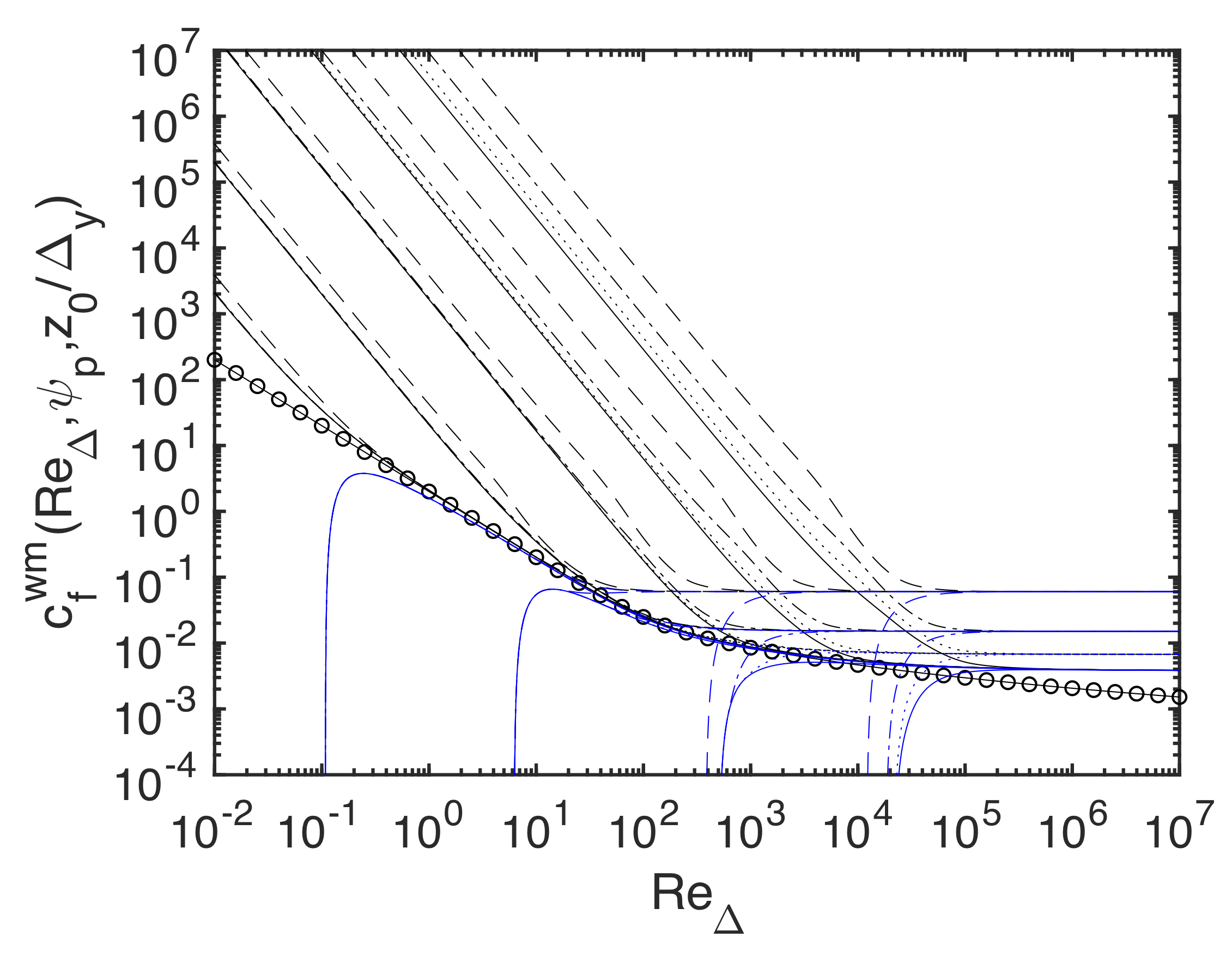}
    \caption{Generalized fitted Moody diagram for wall modeled LES combining smooth and rough wall results, including strong pressure gradient and near-separation regimes. Line types same as in Fig. \ref{fig:retauvsredpsiroughall} .}
    \label{fig:moodystrongall}
\end{figure}

The same results as in Fig. \ref{fig:retauvsredpsiroughall}(b) are presented in the more familiar `friction factor' form, by plotting the corresponding $c_{\rm f}^{\rm wm}=2(Re_{\tau \Delta}^{\rm ufs}/Re_{\Delta})^2$ as function of $Re_\Delta$ for various values of $\psi_p$ and $z_0/\Delta_y$. The familiar bend towards fully rough horizontal lines at high $Re_\Delta$ is again visible. The favorable pressure gradient effects are noteworthy, with a steep $Re_\Delta^{-2}$ scaling at lower $Re_\Delta$ values, caused by a $Re_\Delta$-independent asymptote for $Re_{\tau \Delta}$ at large favorable pressure gradient. In that limit, the expected $c_{\rm f}^{\rm wm} \sim |\psi_p|^{1/2}$ can also be observed. 
 
 Finally, for convenience, we here reproduce the equations required in practice to implement the wall model fit presented in this section.
With inputs $U_{\rm LES}$, $\Delta_y$ and fluid viscosity $\nu$,  
evaluate  
$ Re_\Delta = \frac{U_{\rm LES} \Delta_y} {\nu}.$
For the simplest applications (no pressure gradient, no roughness), Eq. \ref{eq:firstfit} for $Re_{\tau \Delta}^{\rm fit}$  then provides the baseline version of the wall model. 
For inclusion of pressure gradients, using  $\rho^{-1} \partial p_{\rm LES}/\partial s$, $\Delta_y$ and viscosity, evaluate
$
\psi_p = \frac{1}{\rho} \frac{\partial p_{\rm LES}}{\partial s} \, \frac{\Delta^3}{\nu^2}
$
and the friction Reynolds number is obtained as $Re_{\tau\Delta}^{\rm pres}$ using Eqs. \ref{eq:fitretaupresspsineg} and \ref{eq:fitretaupresspsipos} for smooth surfaces and from 
Eq. \ref{eq:ufsfit} when combined to include the rough surface case.
In schematic form, the expressions are:
\begin{align}
 \underline{  {\rm FUNCTION} \,\,\,  Re_{\tau\Delta}^{\rm ufs} }  & \underline{ = Re_{\tau\Delta}^{\rm ufs}(Re_\Delta,\psi_p,z_0/\Delta)   }   \nonumber  \\
  &   \nonumber  \\
 \text{Check: } \,\,  0< Re_\Delta < 10^7, & \,\,\,\,\,\,\,\,  |\psi_p|<2\times 10^7, \,\,\,\,\,\,\,\,  10^{-5}<z_0/\Delta_y < 0.1 .  \nonumber\\
  \nonumber\\
\beta_1 \,\,\,\,   =[{1+0.155/Re_\Delta^{0.03}}]^{-1} & \,\,, \,\,\,\,\,\,  
  \beta_2=1.7-[ {1+36/Re_\Delta^{0.75}} ]^{-1} \,\,, \,\,\,\,
 \nonumber \\
\kappa \,\,\,\,\,\,   =  0.40\,\, , \,\,\,\,\,\,
\kappa_3 & =0.005 \,\, , \,\,\,\,\,\, \kappa_4=\kappa_3^{\beta_1-1/2}\,\,  ,\nonumber \\
 Re^{\rm fit}_{\tau \Delta} & =   \kappa_4 \, Re_\Delta^{\beta_1} \, [1+(\kappa_3 Re_\Delta)^{-\beta_2} ]^{(\beta_1-1/2)/\beta_2}.    \nonumber \\
 \nonumber  \\
\text{To include pressure }& \text{gradients (smooth surface):}  \nonumber  \\
 \text{For }  \psi_p<0:~~~~~~~ Re_{\tau\Delta-{\rm min}}(\psi_p) & =1.5 \, (-\psi_p)^{0.39} \, \left[1+\left(\frac{1000}{(-\psi_p)}\right)^2 \right]^{-0.055} \, ,\nonumber \\
 p(\psi_p) & = 2.5-0.6\left[1+\tanh(2(\log_{10}(-\psi_p)-6))\right], \, \nonumber \\
 Re_{\tau \Delta}^{\rm pres} & = \left( (Re_{\tau\Delta-{\rm min}}(\psi_p) )^{p(\psi_p)} + (Re_{\tau\Delta}^{\rm fit})^{p(\psi_p)} \right)^{1/p(\psi_p)}.\,   \nonumber \\
\text{For }  \psi_p>0 :  ~~ Re_{\Delta-{\rm min}}(\psi_p) & =  2.5 \, \psi_p^{0.54} \, \left(1+\left[\frac{30}{\psi_p}\right]^{1/2} \right)^{-0.88} \, , \nonumber  \\
\text{For }  Re_\Delta   > Re_{\Delta-{\rm min}}: & 
 Re_{\tau \Delta}^{\rm pres}   =  Re_{\tau\Delta}^{\rm fit} \left(1-\frac{1}{(1+\log[Re_\Delta/Re_{\Delta-{\rm min}}(\psi_p)])^{1.9}} \right).\, \nonumber \\
 \text{For } Re_\Delta  \leq Re_{\Delta-{\rm min}}: &
  Re_{\tau \Delta}^{\rm pres}   = 0.  \nonumber  \\
   & \nonumber \\
 \text{To merge with }& \text{ rough-wall representations:}   \nonumber  \\
 \Psi_p   = \psi_p / Re_\Delta^2, \,\,\,\xi = &  [\log(\Delta_y/z_0)]^{-1}, \,\,\,
 \alpha_\Psi    = 1.15 \, |\Psi_p|^{1/2}, \nonumber  \\
  &  \nonumber  \\
\text{For} \,\, \Psi_p<0: & \,\, C(\Psi_p) = 0.085 \,|\Psi_p|^{1/2},  \,\,\, \xi_m = 0.95\, |\Psi_p|^{1/2} \nonumber \\
\text{For} \,\, \Psi_p>0: & \,\, C(\Psi_p) = -0.63\, \Psi_p^{1.24},\,\,\,\,\,\, \xi_m = 0.20 \nonumber \\
 \theta_1(z_0/\Delta_y,\Psi_p) = & - {\rm sign}(\Psi_p)\,
 \sqrt{|\Psi_p|} \,\, \alpha_\Psi\, \left[1+\left( \frac{2.25 \, \xi}{\alpha_\Psi}     \right)^{-1.35} \right]^{-1/1.35}, \nonumber \\
 \theta_2(z_0/\Delta_y,\Psi_p) = &\,C(\Psi_p) \, \frac{\xi}{\xi_m}\, \exp\left(\frac{1}{2}\left[1-\left(\frac{\xi}{\xi_m}\right)^2 \right] \right), \nonumber \\
\Theta^{\rm fit}(z_0/\Delta_y,\Psi_p) = & \, \kappa \, \xi \, + \theta_1(z_0/\Delta_y,\Psi_p) \, + \, \theta_2(z_0/\Delta_y,\Psi_p).\nonumber \\
Re_{\tau\Delta}^{\rm ufs} & = \, \left[  \left( Re_{\tau \Delta}^{\rm pres}   \right)^6  \,+\, \left( Re_{\Delta} \,\,\Theta^{\rm fit}   \right)^6 \right]^{1/6}. \label{eq:table}
\end{align}
The friction velocity can then be obtained as 
\be
u_\tau = U_{\rm LES}\, \frac{Re_{\tau\Delta}^{\rm ufs}}{Re_\Delta}.
\ee

This `universal' fit contains all of the special cases discussed for general (possibly strong) pressure gradient and can thus be implemented without having to establish conditions ahead of time. For mild pressure gradients, the fit for $Re_{\tau\Delta}^{\rm uf}$ (Eq. \ref{eq:univfit}) can be considered of higher accuracy in representing the RANS solutions for small $|\chi|$ (a limit $|\chi|<0.2$ may be considered), see Appendix B, although the effect of $\chi$ on $Re_{\tau\Delta}$ is rather small for $|\chi|<<1$ anyhow.
 
\section{Conclusions}

The main results of this note are the baseline fit of Eq. \ref{eq:firstfit} and the comprehensive fit for general pressure gradients and including roughness, in Eqs. \ref{eq:ufsfit}. For convenience the entire set of fitting functions proposed herein are summarized in Eq. \ref{eq:table}. These fitting functions enable efficient evaluation of friction velocity and wall stress in WMLES, unifying smooth wall and rough wall behaviors, including effects of arbitrary pressure gradients, as well as smoothly merging towards the viscous sublayer. Also, fits are provided for mild pressure gradients (summarized in Appendix B) that can be considered more accurate (maximum relative errors below 2\%), while those for general pressure gradients involve larger errors (possibly up to 4-8\% in some very strong pressure gradient cases with large roughness, for which the underlying RANS model is expected to be inaccurate anyhow). It is important to recall that the fits proposed herein are based on classic mixing length RANS modeling to determine the mean velocity profile from the simplified boundary layer momentum equation, thus inheriting the drawbacks associated with the various underlying assumptions. 

In order to highlight the subtleties and possible differences due to limitations of the standard mixing length RANS model, a further alternate fit is provided in Appendix A based on the empirical wall model of Ref. \cite{nickels2004inner} in which the turbulence structure is known to be directly affected by pressure gradient. In some parameter ranges and for mild pressure gradients, the model (and fitted wall function) corresponds to a non-intuitive effect on the relationship between near wall velocities and wall stress.  As discussed in Ref. \cite{russo2016linear} there even exists the possibility that such subtle trends in near-wall regions arising from imposed body forces (such as $N$ here) cannot be described by any local eddy-viscosity closure. 

The fits are proposed here to facilitate and unify implementation in LES codes. However, we believe that explicit expressions not only assist LES implementation but also help to better understand the relevance of various physical effects and asymptotic limits. 
When using a classic wall-functions approach in WMLES these effects are typically hidden from view since the models are based on assumed velocity profiles that need to be numerically inverted (or they are numerically obtained by integrating an ODE) on the fly. Viewing the entirety of the regimes explicitly as presented here (e.g. in Fig. \ref{fig:moodystrongall}), based on a dimensionless treatment that uses only known parameters for scaling (and avoids the unknown friction velocity) provides, we believe, an interesting perspective to classical wall modeling. 

\section*{Acknowledgments}
The author thanks M. Fowler, Y. Hue, G. Narasimhan, T. Zaki, P. Luchini, J. Larsson and X.I.A. Yang for insightful conversations and comments. Financial support for the present work was provided by the Office of Naval Research (grant \# N00014-17-1-2937) and the National Science Foundation (grant \# CBET-1738918). 

\section*{Appendix A: Alternate fit for Nickels' model at mild pressure gradients}
\label{sec-nickels}
As noted by Nickels (2004) \cite{nickels2004inner} when examining data from FPG   \cite{spalart1986numerical} and APG \cite{nagano1993effects} boundary layers the effect in the inner portion appears to be opposite of predictions of the eddy-viscosity model. Specifically, data seem to suggest a  downward shift for APG and upward shift for FPG. This line of study was further pursued in Refs. \cite{johnstone2009resilience,coleman2015direct} who noted the surprising resilience of the standard log-law from being modified due to pressure gradients. As a synthesis, Nickels proposed a generalized law of the wall including pressure gradient effects that also uses a wall distance $y_c$ representing the pressure-gradient dependent height from the wall where the viscous and turbulent layers transition. In this model, the pressure gradient is represented by the parameter 
\be
p_x^+ = \frac{N \nu}{u_\tau^3} = \chi \, Re_{\tau\Delta}^{-1}.
\ee
Nickels' model for the velocity profile reads
\be
u^+(y) = y_c^+ \left[1-\left(1+2z+0.5(3-p_x^+y_c^+)z^2-1.5p_x^+y_c^+ z^3\right) e^{-3z} \right]+\frac{\sqrt{1+p_x^+y_c^+}}{6 \kappa} \log[1+(a z)^6],
\ee
where $z=y/y_c$, $a=0.75$ (Nickels used $a=0.6$ but 0.75  has to be used for consistency with $\kappa = 0.4$ and $B=5$ used elsewhere in this paper). Above, $y_c^+$ is obtained from solving
\be
p_x^+ {y_c^+}^3 + {y_c^+}^2 = 12^2.
\label{eq-ycp}
\ee
For present purposes, the solution to Eq. \ref{eq-ycp} can be fitted using a $\tanh$ function to capture the main behavior near $p_x^+ \sim 0$ but avoiding unphysical divergences when $p_x^+$ is large: 
\be 
y_c^+ = 12 - 5 \,\tanh(12.4 \, p_x^+).
\ee

\begin{figure}[h]
    \centering
    \includegraphics[width=0.70\textwidth]{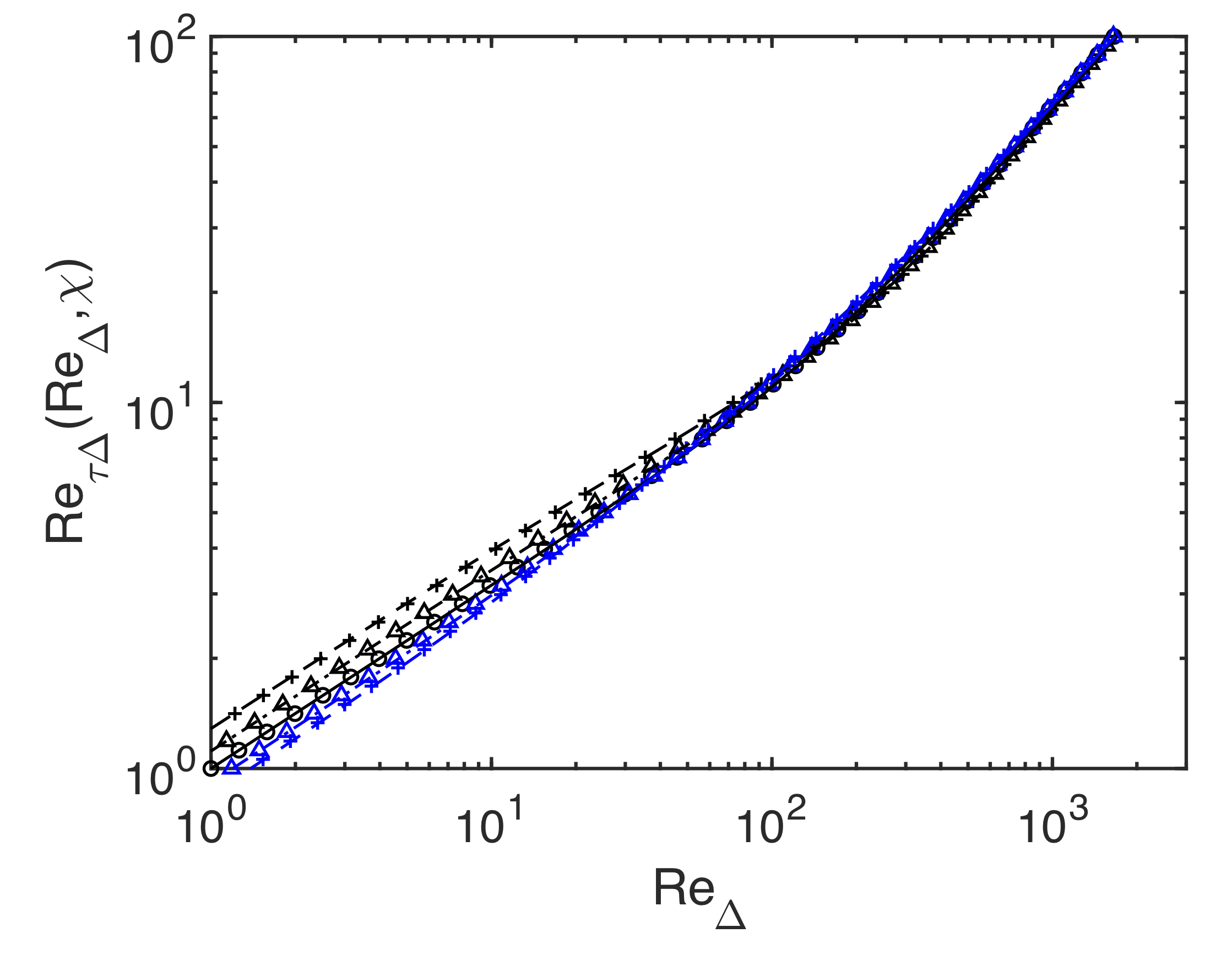}
    \caption{Symbols: Results from Eq. \ref{eq:ReDeltaNick} over a range of Reynolds numbers $Re_{\tau \Delta}$, for $\chi$ = -0.8 (black +), $\chi$ = -0.4 (black triangles),  $\chi$=0 (circles), $\chi$ = 0.4 (blue triangles), $\chi$ = 0.8 (blue +).  Only the region between $1<Re_\Delta<3000$ is shown for clarity. Lines: empirical fit given by Eq. \ref{eq:RetauDeltaNick}. Solid line: $\chi=0$, dot-dashed lines: 
    $|\chi|=0.4$, dashed line: $|\chi|=0.8$. Black: favorable pressure gradient ($\chi \leq 0$), blue lines: adverse pressure gradient $\chi>0$. Recall that $p_x^+ = \chi / Re_{\tau\Delta}$.}
    \label{fig-retvsredxNick}
\end{figure}

\begin{figure}
        \centering
        \subfloat[\label{fig:fig2}]{\includegraphics[width=0.5\columnwidth,trim=4 4 4 4,clip]{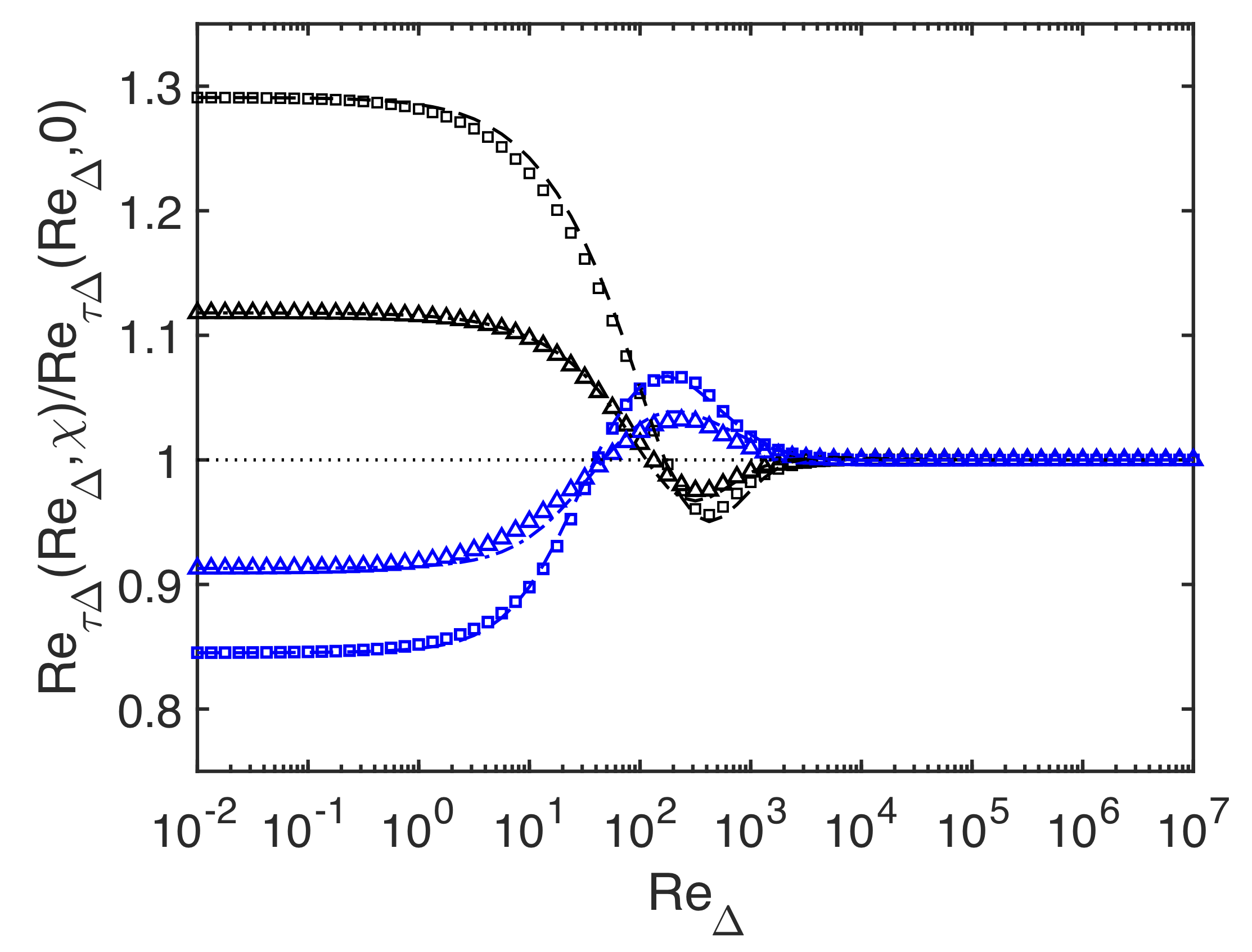}}
        \hfill
       \subfloat[\label{fig:dsd_3_5}] {\includegraphics[width=0.5\columnwidth,trim=4 4 4 4,clip]{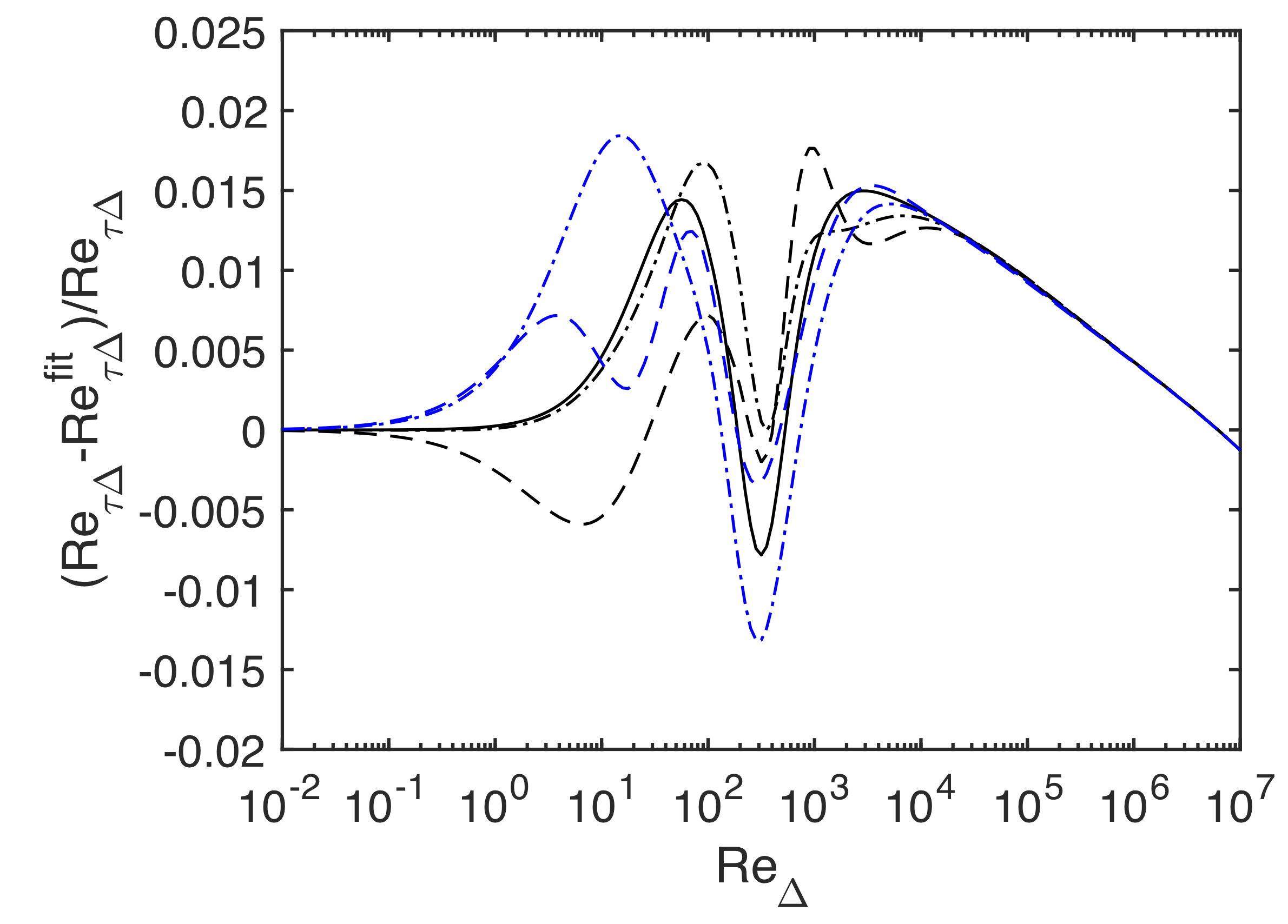}}
  \caption{ 
  (a) Symbols: Ratio of friction Reynolds number as function $Re_\Delta$ from Nickels' model, for:  $\chi$ = -0.8 (black squares), $\chi$ = -0.4 (black triangles), 
    $\chi$ = 0.4 (blue triangles), $\chi$ = 0.8 (blue squares).  The relative effect of pressure gradient is larger at lower Reynolds number.  The lines are from an empirical fit (Eq. \ref{eq:RetauDeltaNick}). (b)  Relative error between the Nickels model (inverse of Eq. \ref{eq:ReDeltaNick}) and empirical fit given by Eq. \ref{eq:RetauDeltaNick}. Solid line: $\chi=0$, dot-dashed lines: $|\chi|=0.4$, dashed line: $|\chi|=0.8$. Black: favorable and zero pressure gradient  cases($\chi \leq 0$), blue lines: adverse pressure gradient $\chi>0$.
  } 
    \label{fig-ratioerrorNick}
\end{figure}

Using our nomenclature related to wall modeling, the above expression evaluated at $y=\Delta_y$ can be written as
$$
Re_{\Delta} = Re_{\tau \Delta} \, \, y_c^+ \,  \left[1-\left(1+2z+0.5(3-p_x^+y_c^+)z^2-1.5p_x^+y_c^+z^3\right) e^{-3z} \right] + ...
$$
\be
\hskip 1in  .. + \, \frac{\sqrt{1+p_x^+y_c^+}}{6 \kappa} \log[1+(0.75 z)^6],
\label{eq:ReDeltaNick}
\ee
with $z=Re_{\tau \Delta}/y_c^+$. For a range of $\chi$ and $Re_{\tau\Delta}$ values, we evaluate $Re_\Delta$ and plot $Re_{\tau\Delta}$ as function of $Re_\Delta$ in Fig \ref{fig-retvsredxNick}. The result is similar to Fig. \ref{fig4} but for the middle region where some non-monotonic behavior can be seen where curves cross. The differences can be appreciated more clearly in Fig. \ref{fig-ratioerrorNick}(a) where the ratio to the case with $\chi=0$ is shown. Near $Re_\Delta \sim 200$ the behavior is non-monotonic and there exists a region in which indeed APG yields larger wall stress ($Re_{\tau \Delta}$) for a given velocity ($Re_{\Delta}$) and FPG a lower one, as described by the fit proposed by Nickels and consistent with data from Refs. \cite{spalart1986numerical,nagano1993effects}.

Aiming to fit these results, we propose slight modifications to the basic fit provided in Eq. \ref{eq:firstfit}. Since the asymptotic viscous and log-layer regimes are the same, only the transition behavior parameter $\beta_2$ is modified slightly for the ZPG ($\chi=0$) baseline case used to evaluate $Re_{\tau \Delta}^{\rm fit}(Re_{\Delta})$: 
  \be
  \beta_2(Re_{\Delta}) = 1.7-\left(1+36 Re_{\Delta}^{-0.65}\right)^{-1}.
 \label{eq:beta2Nick}
 \ee
Including the Nickels model for pressure gradients, the final form of the proposed fit reads:
 \be 
 Re_{\tau \Delta}^{\rm nic} = Re_{\tau \Delta}^{\rm fit}(Re_{\Delta})\left[ \theta (1+\chi/2)^{-1/2} + 1 - \theta + \gamma(Re_{\Delta},\chi) \right],
 \label{eq:RetauDeltaNick}
 \ee
 where $\theta$ is given again by $\theta(Re_{\Delta}) = (1+0.0025 Re_{\Delta})^{-1}$, and
 \be
 \gamma(Re_{\Delta},\chi) = \alpha(\chi) \exp\left[-\frac{(\log_{10} Re_\Delta - \mu(\chi))^2}{2\sigma^2(\chi)} \right],
 \ee
 and the other parameters are fitted to avoid unphysical limits at large $\chi$:
$\alpha(\chi) = 0.0296 + 0.15 \tanh(\chi-0.2)$, $\mu(\chi) =  2.25 -0.4 \tanh(0.9\chi)$,  and $\sigma(\chi) = 0.5 + 0.1 \tanh(\chi/0.05)$. 
The solid lines in Figs. \ref{fig-retvsredxNick} and \ref{fig-ratioerrorNick}(a) show the resulting fits. The relative errors are shown in Fig. \ref{fig-ratioerrorNick}(b), falling below 2\%. Hence, for mild pressure gradient applications in which the subtle pressure gradient effects as described in Refs. \cite{nickels2004inner,johnstone2009resilience} are to be included instead of those arising from the standard eddy-viscosity assumption, $Re_\Delta^{\rm nic}$ can be used instead of $Re_\Delta^{\rm com}$.

\section*{Appendix B: Summary of fitting function for mild pressure gradient}

Returning to the standard eddy-viscosity approach discussed in \S \ref{sec-mild}, for inclusion of mild pressure gradient without roughness,  and using    $N=\rho^{-1} \partial p_{\rm LES}/\partial s$, $\Delta_y$ and the baseline friction velocity from \ref{eq:firstfit}, evaluate $\chi$ 
from
$$
\chi =   \frac{N \Delta_y} {U_{\rm LES}^2} \,\, 
\left(\frac{Re_\Delta} {Re_{\tau \Delta}^{\rm fit}} \right)^2
$$
and then the combined model $Re_{\tau \Delta}^{\rm com}$ according to Eq. \ref{eq6} provides the model outcome.
For inclusion of roughness in the fully rough regime, one would evaluate $\chi$ according to
$$
\chi = \frac{N \Delta_y} {U_{\rm LES}^2} \,\, \left( \frac{1}{\kappa} \, \log(\Delta_y/z_0) \right)^2
$$
and evaluate the friction Reynolds number according to Eq. 
\ref{eq:reyninf}. 
For inclusion of roughness as well as mild pressure gradients and viscous effects, evaluate $\chi$ using Eq. \ref{eq:chipractical}, rewritten as
$$
\chi = \frac{N \Delta_y} {U_{\rm LES}^2} \,\,\, \min\left[
\frac{Re_\Delta} {Re_{\tau \Delta}^{\rm fit}}\, , \,\, \frac{1}{\kappa} \, \log(\Delta_y/z_0) \right]^2.
$$
To ensure validity of the fits and derivations, in practice $\chi$ has to be small, e.g. to fall between $-0.2$ and +0.2, i.e.
use ${\rm sign}(\chi) \min(|\chi|,0.2)$. Then, one determines $Re_{\tau\Delta}^{\rm uf}$ from Eq. \ref{eq:univfit}.
 
\begin{align}
 \underline{ {\rm FUNCTION} \,\,\,  Re_{\tau\Delta}^{\rm uf} }   & \underline{ = Re_{\tau\Delta}^{\rm uf}(Re_\Delta,\chi,z_0/\Delta)}   \nonumber \\
 \nonumber  \\
 \text{Check: } \,\, & 0< Re_\Delta < 10^7, \,\,\,\, |\chi|<0.2, \,\,\, 0<z_0/\Delta_y < 0.1 .  \nonumber\\
  &  \nonumber\\
\beta_1 \,\,\,\,& =[{1+0.155/Re_\Delta^{0.03}}]^{-1} \,\,, \,\,\,\,\,\,
\beta_2=1.7-[ {1+36/Re_\Delta^{0.75}} ]^{-1} \,\,, \,\,\,\,
 \nonumber \\
\kappa \,\,\,\,\,\, & =  0.40\,\, , \,\,\,\,\,\,
\kappa_3 =0.005 \,\, , \,\,\,\,\,\, \kappa_4=\kappa_3^{\beta_1-1/2}\,\,  ,\nonumber \\
 Re^{\rm fit}_{\tau \Delta} & =   \kappa_4 \, Re_\Delta^{\beta_1} \, [1+(\kappa_3 Re_\Delta)^{-\beta_2} ]^{(\beta_1-1/2)/\beta_2}.    \\
 \nonumber  \\
\text{To include }& \text{mild pressure gradients:}  \nonumber  \\
 Re_{\tau \Delta,v} & = (1+0.5\, \chi)^{-1/2} \, Re^{\rm fit}_{\tau \Delta} \, ,\nonumber \\
 Re_\Delta^*&  =  Re_\Delta - \frac{\chi}{2\kappa}\, Re^{\rm fit}_{\tau \Delta} \, (1-11/Re^{\rm fit}_{\tau \Delta})\, [1+(50/Re^{\rm fit}_{\tau \Delta})^{2}]^{-1/2} \, , \nonumber \\
\beta^*_1 \,\,\,\,  &=[{1+0.155/{Re^*_\Delta}^{0.03}} ]^{-1}, \,\,\,\,\,\,
\beta^*_2=1.7- [{1+36/{Re^*_\Delta}^{0.75}} ]^{-1} \, , \nonumber \\
\kappa^*_4\,\,\,\,&= \kappa_3^{\beta^*_1-1/2}\, , \nonumber \\
Re_{\tau \Delta,{\rm in}} & =  \kappa^*_4 \, ({Re^*_\Delta})^{\beta^*_1} \,  [ 1+(\kappa_3 {Re^*_\Delta})^{-\beta^*_2} ]^{(\beta^*_1-1/2)/\beta^*_2} \, ,\nonumber \\
\theta \,\,\, & = \left(1+Re_\Delta/400 \right)^{-1} \, ,   \nonumber \\
Re_{\tau \Delta}^{\rm com} & = \theta \,\, Re_{\tau \Delta,v} \,\,+\,\, (1-\theta) \,\, Re_{\tau \Delta,{\rm in}} .   \\
& \nonumber  \\
 \text{To merge with  }& \text{ rough-wall representations:}   \nonumber  \\
Re_{\tau \Delta}^{\infty} & = Re_\Delta \,\, \left[ \frac{1}{\kappa} \log(\Delta/z_0) + \frac{\chi_{x}}{2 \kappa} (1-{z_0}/{\Delta}) \right]^{-1} \, ,   \\
Re_{\tau\Delta}^{\rm uf} & = \, \left[  \left( Re_{\tau \Delta}^{\rm com}  \right)^6  \,+\, \left( Re_{\tau \Delta}^{\infty}  \right)^6 \right]^{1/6}. 
\end{align}

\noindent For applications of the Nickels model discussed in Appendix A, $Re_{\tau\Delta}^{\rm com}$ above can be replaced by 
$Re_{\tau\Delta}^{\rm nic}$ (Eq. \ref{eq:RetauDeltaNick}). 

\newpage 
\bibliography{meneveau-biblio}
\end{document}